\documentclass{article}
\usepackage[12pt]{extsizes}
\usepackage[utf8]{inputenc}
\usepackage[sort]{natbib}
\usepackage{graphicx}
\usepackage{lipsum}

\usepackage{enumitem}
\usepackage{mydef}
\usepackage{caption}
\usepackage{mdframed}
\usepackage[hidelinks]{hyperref}
\definecolor{myurlcolor}{HTML}{123463}
\definecolor{tdc_color}{RGB}{10,128,122}
\definecolor{dc_color}{RGB}{230, 245, 244}
\definecolor{ds_color}{RGB}{195, 230, 227}
\definecolor{ms_color}{RGB}{150, 214, 209}
\hypersetup{
    colorlinks=true,
    linkcolor=tdc_color,
    filecolor=tdc_color,      
    urlcolor=tdc_color,
    citecolor=tdc_color
}
\usepackage{authblk}

\usepackage{geometry}
\usepackage{relsize}
\usepackage{multirow}
\usepackage{booktabs}
\usepackage{tabularx}
\usepackage{bbding}
\usepackage{listings}
\usepackage{xspace}
\usepackage[ruled,vlined]{algorithm2e}
\definecolor{light-gray}{gray}{0.96}

\usepackage[p,osf]{cochineal}
\usepackage[scale=.95,type1]{cabin}
\usepackage[zerostyle=c,scaled=.94]{newtxtt}
\usepackage{microtype}
\usepackage{subfig} 
\usepackage{subfloat}
\usepackage{adjustbox}
\usepackage{etoolbox}
\apptocmd{\thebibliography}{\raggedright}{}{}
\usepackage{lmodern}

\usepackage{dirtree}
\makeatletter
\patchcmd{\@maketitle}{\LARGE \@title}{\huge\@title}{}{}
\makeatother
\usepackage{threeparttable}

\title{\textbf{Protein Language Models and Structure Prediction: Connection and Progression} \\[1ex] {\LARGE{\textbf{A Systematic Survey}}}}
\author[$\dagger$, $\ddagger$]{Bozhen Hu}
\author[$\ddagger$]{Jun Xia}
\author[$\ddagger$]{Jiangbin Zheng}
\author[$\ddagger$]{Cheng Tan}
\author[$\ddagger$]{Yufei Huang}
\author[$\ddagger$]{Yongjie Xu}
\author[$\ddagger$]{Stan Z. Li}
\affil[$\dagger$]{Zhejiang University, Hangzhou, 310058, China}
\affil[$\ddagger$]{AI Division, School of Engineering, Westlake University, Hangzhou, 310030, China}
\affil[ ]{}
\affil[ ]{\texttt{hubozhen;xiajun;zhengjiangbin;}}
\affil[ ]{\texttt{tancheng;huangyufei;xuyongjie@westlake.edu.cn}}
\affil[ ]{Correspondence: ~\href{mailto:Stan.ZQ.Li@westlake.edu.cn}{\color{tdc_color}\texttt{Stan.ZQ.Li@westlake.edu.cn}}}
\date{}
\begin{document}
\maketitle
\begin{abstract}
    \normalsize
\noindent The prediction of protein structures from sequences is an important task for function prediction, drug design, and related biological processes understanding. Recent advances have proved the power of language models (LMs) in processing the protein sequence databases, which inherit the advantages of attention networks and capture useful information in learning representations for proteins. The past two years have witnessed remarkable success in tertiary protein structure prediction (PSP), including evolution-based and single-sequence-based PSP. It seems that instead of using energy-based models and sampling procedures, protein language model (pLM)-based pipelines have emerged as mainstream paradigms in PSP. Despite the fruitful progress, the PSP community needs a systematic and up-to-date survey to help bridge the gap between LMs in the natural language processing (NLP) and PSP domains and introduce their methodologies, advancements and practical applications. To this end, in this paper, we first introduce the similarities between protein and human languages that allow LMs extended to pLMs, and applied to protein databases. Then, we systematically review recent advances in LMs and pLMs from the perspectives of network architectures, pre-training strategies, applications, and commonly-used protein databases. Next, different types of methods for PSP are discussed, particularly how the pLM-based architectures function in the process of protein folding. Finally, we identify challenges faced by the PSP community and foresee promising research directions along with the advances of pLMs. This survey aims to be a hands-on guide for researchers to understand PSP methods, develop pLMs and tackle challenging problems in this field for practical purposes.
\end{abstract}

\clearpage
{\normalsize
\tableofcontents
\listoftables
}
\clearpage
\section{Backgrounds}
Proteins are the workhorses of life, playing an essential role in a broad range of applications ranging from therapeutics to materials. They are built from 20 different basic chemical building blocks (called amino acids), which fold into complex ensembles of 3-dimensional structures that determine their functions and orchestrate the biological processes of cells. However, predicting protein structure from amino acid sequence is challenging because small perturbations in the sequence of a protein can drastically change the protein's shape and even render it useless, while different amino acids can have similar chemical properties, so some mutations will hardly change the shape of the protein. What is worse, the polypeptide is flexible and can fold into a staggering number of different shapes ~\citep{kryshtafovych2019critical,senior2020improved}. Thus, PSP has long been a central but incompletely tackled problem in the scientific community. 

One way to find out the structure of a protein is to use an experimental approach, including X-ray crystallography, Nuclear Magnetic Resonance (NMR) Spectroscopy~\citep{ikeya2019protein} and cryo-electron microscopy (cryo-EM)~\citep{gauto2019integrated}. The experimental protein structures have been deposited in the Protein Data Bank (PDB)~\citep{wwpdb2019protein}. Unfortunately, laboratory approaches for structure determination are expensive and cannot be used on all proteins.
This challenge makes the number of reported protein structures orders of magnitude lower than the size of datasets in other machine-learning application domains. For example, 190 thousand structures exist in PDB~\citep{HelenMBerman2000ThePD} versus 528 million protein sequences in UniParc~\citep{2012UpdateOA} and versus 10 million annotated images in ImageNet~\citep{OlgaRussakovsky2014ImageNetLS}.

In general, computational methods for predicting three-dimensional (3D) protein structures from protein sequences have traditionally taken two parallel paths, focusing on either physical interactions or evolutionary principles~\citep{JohnMJumper2021HighlyAP}. Since proteins generally fold into their lowest free energy states, the physics-based approach simulates the folding process of the amino acid chain using molecular dynamics based on the potential energy of the force field at a particular time or fragment assembly using the energy function, which concentrates on the physical interactions to form an energy-stable 3D structure. However, this approach has proved highly challenging for even moderately sized proteins due to the computational intractability of molecular simulation, the conditioned accuracy of fragment assembly, and the difficulty of producing sufficiently accurate models of protein physics~\citep{alquraishi2019end, susanty2021review}. On the other hand, thanks to the recent progress in protein sequencing~\citep{BinMa2012DeNS, BinMa2015NovorRP}, a large number of protein sequences are now available. For example, the UniProt~\citep{AlexBateman2019UniProtAW} database contains over 200 million protein sequences with relevant information. These protein databases support to get multiple sequence alignment (MSA) of homologous proteins, which significantly benefits the development of evolutionary methods.  

\begin{figure*}[t]
	\begin{center}
		\includegraphics[width=0.92\linewidth]{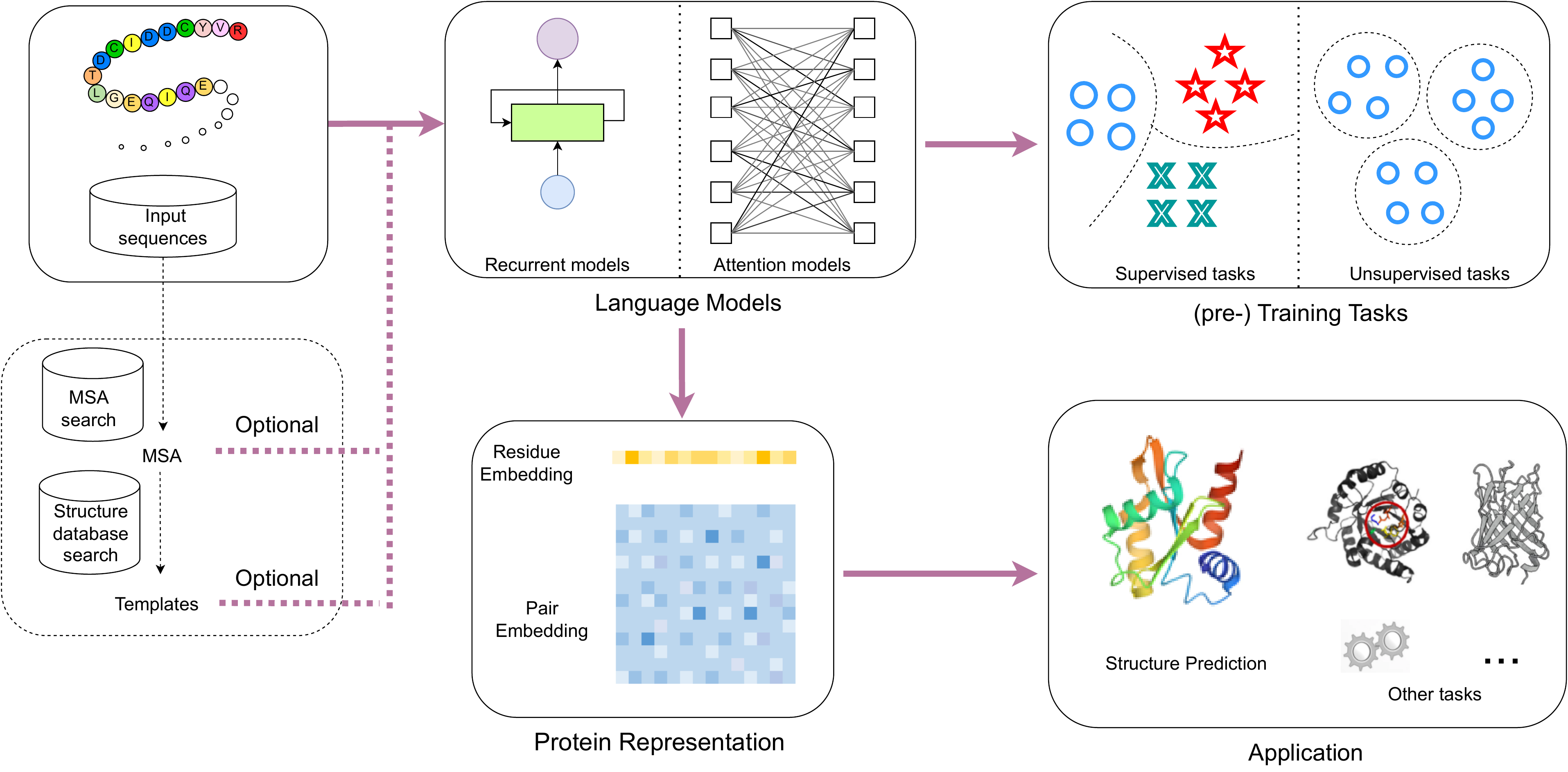}
	\end{center}
	\caption{Schematic diagram of LMs for learning protein representations  for PSP. Protein structures can be leveraged as labels in supervised tasks; dashed arrows represent optional.}
	\label{fig_01}
\end{figure*}

LMs have recently emerged as a powerful paradigm for learning "content-aware" data representations from large-scale sequence databases~\citep{TristanBepler2021LearningTP}, which are widely used for machine translation, question answering in NLP~\citep{JacobAndreas2013SemanticPA} and are even extended to computer vision~\citep{VuPham2013DropoutIR}, molecules~\citep{xia2022pre}, etc. Due to the similarities between protein and human languages, LMs are gradually modified into pLMs to deal with various protein data, specially matched for protein sequences to learn representations that can be used for PSP. Large-scale pLMs with self-supervised pre-training on tens of millions to billions of proteins~\citep{RoshanRao2019EvaluatingPT, AlexanderRives2019BiologicalSA, AhmedElnaggar2022ProtTransTC, TristanBepler2022LearningTP} are the current state-of-the-art methods in predicting protein structure, function, and fitness from sequences. For example, the introductions of MSA input and pLMs have led to the vast success of AlphaFold2 (AF2)~\citep{JohnMJumper2021HighlyAP} at the Critical Assessment of protein Structure Prediction (CASP) 14 competition~\citep{kryshtafovych2019critical}. Since the propose of AF2 and RosettaFold ~\citep{MinkyungBaek2021AccuratePO}, discussions on PSP and related protein tasks have culminated, and more researchers have begun to develop pLMs and tackle challenging problems that have not yet been solved, including PSP without evolutionary information by the usage of pLMs, making accurate structure predictions for protein complexes~\citep{RichardEvans2021ProteinCP}, finding mechanisms behind protein folding, etc. Figure~\ref{fig_01} shows the overall pipeline of pLMs for the prediction of protein 3D structure.

Protein representation learning, inspired by approaches in NLP, is an active area of research that learns representations used for various downstream tasks~\citep{unsal2022learning}. However, task-specific labelled proteins can be highly scarce because labelling proteins often requires time-consuming and resource-intensive lab experiments. To relieve and even tackle this problem, a pre-train and then fine-tune paradigm of LMs has been described in NLP. The knowledge is gained through pre-training a model on a source task and is used to improve the learning by fine-tuning the model on a new target task with fewer labels. Fortunately, self-supervised learning is applied to learn protein representation, like the masked language modelling tasks that reconstruct corrupted tokens given the surrounding sequence; well-known pre-trained sequence encoders include TAPE Transformer~\citep{RoshanRao2019EvaluatingPT}, ProteinBert~\citep{DanOfer2021ProteinBERTAU}, ProtTrans~\citep{AhmedElnaggar2021ProtTransTC} and ESM-1b~\citep{AlexanderRives2019BiologicalSA} have been trained by predicting the masked residues in sequences. Besides, GearNet~\citep{ZuobaiZhang2022ProteinRL}, STEPS~\citep{CanChen2022StructureawarePS} are proposed for protein representations to exploit the topology information of structures, which are expected to contain more valuable features.

Although pLMs have been increasingly applied in protein representation learning and PSP, a systematic summary of this fast-growing field is still awaited. The following sections give an explanation of the common-used terms and notions, then summarize similarities between natural languages and proteins, next present commonly-used LMs and pLMs, show their applications, innovations, and differences, and finally introduce how different methods work for PSP, especially for the transformer-based pLMs that are used to reshape the protein representation learning area. 

More importantly, we outline several works for single sequence PSP, structure prediction of antibodies, protein complexes, protein-ligands, protein-RNA, and protein conformational ensembles, etc., that are recently proposed and are future research directions. Besides, traditional physics- and machine-learning-based methods for protein structural feature prediction and PSP, are also presented. Moreover, we collect abundant resources, including LMs and pLMs, methods for PSP, pre-training databases, and paper lists\footnote{\url{https://github.com/bozhenhhu/A-Review-of-pLMs-and-Methods-for-Protein-Structure-Prediction}}. Finally, we provide possible future research directions by discussing the limitations and unsolved problems of existing methods.  

To the best of our knowledge, this is the first survey including pLMs for structure prediction, presenting their connections and developments. We aim to help researchers develop more suitable algorithms and tackle essential, challenging, and urgent problems for proteins that can promote the development of biochemistry, biomedicine, and bioinformatics.

\begin{figure*}[h]
	\begin{center}
		\includegraphics[width=0.95\linewidth]{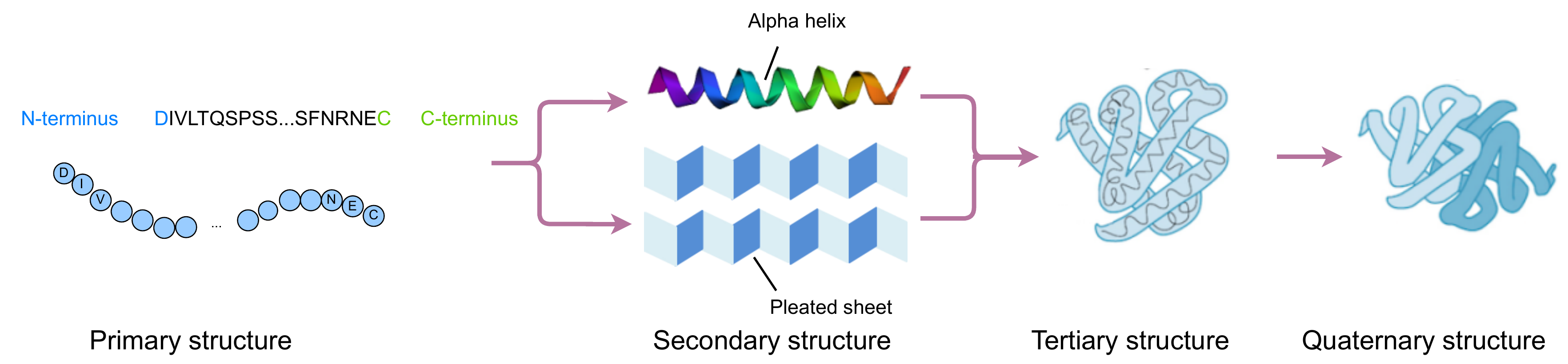}
	\end{center}
	\caption{Four different levels of protein structures~\citep{Ihm2004ATA, Patel2013ProteinSS}.}
	\label{fig_02}
\end{figure*}
\section{Notions and Terms}
\label{Notions and Terms}
\begin{mdframed}[hidealllines=true,backgroundcolor=tdc_color!5]
	\begin{itemize}[leftmargin=*,topsep=0pt]
		\label{terms1}
		\setlength\itemsep{0em}
		\renewcommand{\thempfootnote}{$\star$}
		\item \textbf{Sequence/primary structure}: The linear sequence of amino acids in a peptide or protein~\citep{sanger1952arrangement}. Any sequence of polypeptides is reported starting from the single amine (N-terminus) end to carboxylic acid (C-terminus) ~\citep{hao2017conformational} (Figure~\ref{fig_02}). 
		\item \textbf{Secondary structure (SS)}: The 3D form of local segments of proteins. The two most common secondary structural elements are $\alpha$-helix (H) and $\beta$-strand (E); 3-state SS includes H, E, C (coil region); 8 fine-grained states include three types for helix (G for $3_{10}$-helix, H for $\alpha$-helix, and I for $\pi$-helix), two types for strand (E for $\beta$-strand and B for $\beta$-bridge), and three types for coil (T for $\beta$-turn, S for high curvature loop, and L for irregular)~\citep{ShengWang2015ProteinSS}.
		\item \textbf{Tertiary structure}: The 3D shape of a protein.
		\item \textbf{Quaternary structure}: The 3D arrangement of the subunits in a multisubunit protein~\citep{chou2003predicting}.
		\item \textbf{Multiple sequence alignment (MSA)}: The result of the alignment of three or more biological sequences (protein or nucleic acid).
		\item \textbf{Sequence homology}: The biological homology between sequences (proteins or nucleic acids)~\citep{koonin2005orthologs}. MSA assumes all the sequences to be aligned may share recognizable evolutionary homology~\citep{wang2018benchmark} and is used to indicate which regions of each sequence are homologous.
		\item \textbf{Coevolution}: The interdependence between the evolutionary changes of two entities~\citep{ochoa2014practical} plays an important role at all biological levels, which is evident between protein residues (Figure~\ref{fig_03}(a)).		
		
	\end{itemize}
\end{mdframed}

\begin{mdframed}[hidealllines=true,backgroundcolor=tdc_color!5]
	\begin{itemize}[leftmargin=*,topsep=0pt]
		\label{term2}
		\setlength\itemsep{0em}
		\renewcommand{\thempfootnote}{$\star$}
        \item \textbf{Templates}: The homologous 3D structures of proteins.
		\item \textbf{Contact map}: A two-dimensional binary matrix represents the residue-residue contacts of a protein within a distance threshold~\citep{IsaacArnoldEmerson2017ProteinCM}, which produces a reduced representation of a protein structure.
		\item \textbf{Torsion angles}: Two essential torsion angles (dihedral angles) in the polypeptide chain, which describe the rotations of the chain around the bonds between $\text{N}-\text{C}_\alpha$ ($\varphi$) and $\text{C}_\alpha-\text{C}$ ($\psi$), respectively, each involving four atoms.
		
		\item \textbf{Protein structure prediction (PSP)}: The prediction of the 3D structure of a protein from its amino acid sequence.
		\item \textbf{Orphan proteins}: Proteins without any detectable homology~\citep{basile2017high} (MSAs of homologous proteins are not available).
		\item \textbf{Antibody}: A Y-shaped protein is produced by the immune system to detect and neutralize harmful substances, such as viruses and pathogenic bacteria.
		\item \textbf{RNA}: A polymeric molecule essential in various biological roles, most often single-stranded.
		\item \textbf{Protein complex}: A form of quaternary structure associated with two or more polypeptide chains.
		\item \textbf{Protein conformation}: The spatial arrangement of its constituent atoms that determines the overall shape~\citep{JamesCBlackstock1989GuideTB}.
		\item \textbf{Protein energy function}: Proteins fold into 3D structures in a way that leads to a low-energy state. Protein-energy functions are used to guide PSP by minimizing the energy value.
		\item \textbf{Monte Carlo methods}: A class of computational mathematical algorithms that use repeated random sampling to estimate the possible outcomes of an uncertain event.
		\item \textbf{Protein function prediction}: A Task that uses techniques to assign biological or biochemical roles to proteins. Gene Ontology (GO) annotations classify functions into three main categories of molecular function, biological process, and cellular component~\citep{MAshburner2000GeneOT}.
		\item \textbf{Protein stability prediction}: A task that uses methods to predict the impacts of amino acid mutations (substitutions).
		\item \textbf{Protein design}: A technique that design new proteins with novel purpose, behaviour from scratch, known structure or its sequence.
		\item \textbf{Protein structure refinement}: A task that aims to increase the accuracy of starting models (decoys), i.e., closer to their native states.
		\item \textbf{Supervised learning}: The use of labelled input-output pairs to learn a function that can classify data or predict outcomes accurately.
		\item \textbf{Unsupervised learning}: Models are trained without a labelled dataset and encouraged to discover hidden patterns and insights from the given data.
		\item \textbf{Natural language processing (NLP)}: The ability of computer programs to process, analyze, and understand the text and spoken words in much the same way humans can.
		\item \textbf{Language model (LM)}: A probability distribution of words or word sentences.
		\item \textbf{Embedding}: An embedding is a low-dimensional, learned continuous vector representation of discrete variables into which you can translate high-dimensional and real-valued vectors (words or sentences)~\citep{li2016generative}.

	\end{itemize}
\end{mdframed}

\begin{mdframed}[hidealllines=true,backgroundcolor=tdc_color!5]
	\begin{itemize}[leftmargin=*,topsep=0pt]
		\label{term3}
		\setlength\itemsep{0em}
		\renewcommand{\thempfootnote}{$\star$}
		\item \textbf{Convolution Neural Networks (CNNs)}: A class of neural networks that consist of convolutional operations to capture the local information found in the data.
		\item \textbf{Recurrent Neural Networks (RNNs)}: A class of neural networks where connections between nodes form a directed or undirected graph along a temporal sequence.
		\item \textbf{Attention models}: A class of neural networks used to focus on specific components of a complex input and categorize the whole dataset sequentially~\citep{lin2017structured}.
		\item \textbf{Tansfer learning}: A machine learning method where a model developed for one task is reused for a model to solve a different but related task~\citep{KarlRWeiss2016ASO,SinnoJialinPan2010ASO}, which has two major activities, i.e., pre-training and fine-tuning. 
		\item \textbf{Pre-training}: A strategy in AI refers to training a model with one task to help it form parameters that can be used in other tasks.
		\item \textbf{Fine-tuning}: A method that takes the weights of a pre-trained neural network, which are used to initialize a new model being trained on the same domain.
		\item \textbf{Autoregressive language model}: A feed-forward model predicts the future word from a set of words given a context~\citep{SamBondTaylor2021DeepGM}. 
		\item \textbf{Masked language model}: A language model masks some of the words in a sentence and predicts which words should replace those masks.
		\item \textbf{Bidirectional language model}: A language model learns to predict the probability of the next token in the past and future directions~\citep{MuhammadShahJahan2021BidirectionalLM}.
		\item \textbf{Multi-task learning}: A machine learning paradigm in which multiple tasks are solved simultaneously while exploiting commonalities and differences across tasks~\citep{TristanBepler2021LearningTP}.
		\item \textbf{Sequence-to-Sequence (Seq2Seq)}: A family of machine learning approaches train models to convert sequences from one domain to sequences in another domain.
		\item \textbf{Knowledge graph}: A semantic network uses a graph-structured data model or topology to integrate data~\citep{mccusker2018knowledge}.
		\item \textbf{Knowledge distillation}: The process of transferring the knowledge from a large model or set of models to a single smaller model~\citep{JianpingGou2020KnowledgeDA}.
		\item \textbf{Multi-modal learning}: Training models by combining information obtained from more than one modality~\citep{Skocaj2012,wang2022uncertainty}. 
		\item \textbf{Residual neural network}: An artificial neural network (ANN) in which skip connections or shortcuts are used to jump over some layers, e.g., the deep residual network, ResNet~\citep{KaimingHe2016IdentityMI}.
		
	\end{itemize}
\end{mdframed}
\section{Protein and Language}
\label{Protein and Language}
LMs are increasingly applied to large-scale protein sequence databases to learn embeddings for protein structure or function prediction recently~\citep{TomYoung2017RecentTI,KevinKYang2018LearnedPE,EhsaneddinAsgari2015ContinuousDR}. One important reason is that human languages and proteins share common characteristics. Such as the hierarchical organization~\citep{ferruz2022controllable, OFER20211750}, which means that the four different levels of protein structures (see Figure~\ref{fig_02}) analogy to letters, words, sentences, and texts of human languages to a certain degree. It illustrates that proteins and languages typically comprise modular elements that can be reused and rearranged. Moreover, the rules of protein folding, e.g., the hydrophilicity and hydrophobicity of amino acids, the principle of minimal frustration~\citep{bryngelson1995funnels} and the "folding funnel" landscapes of proteins~\citep{leopold1992protein}, etc., are similar to language grammars of linguistics. 
\begin{figure}[t]
	\centering
	\subfloat[Protein]
	{
		\label{fig_03_1}
		\includegraphics[width=0.62\linewidth]{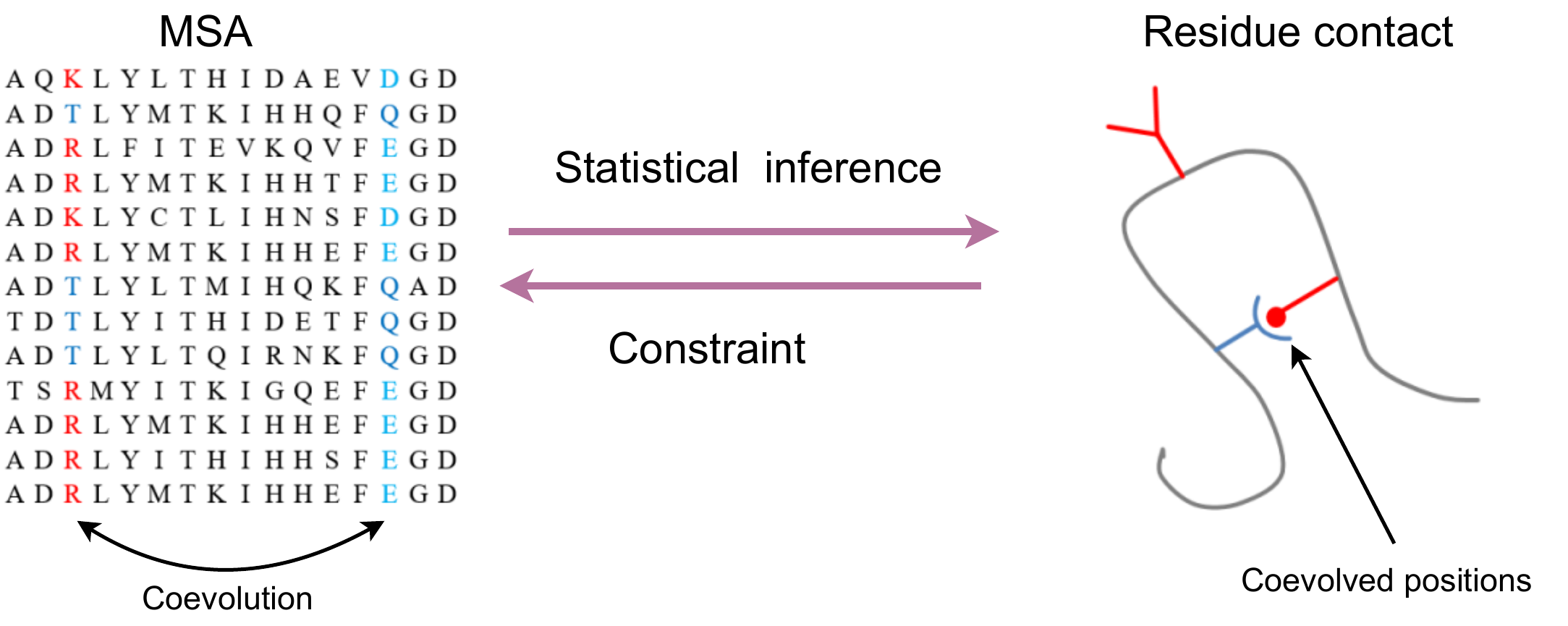}
	}
	\subfloat[Language]
	{
		\label{fig_03_2}
		\includegraphics[width=0.32\linewidth]{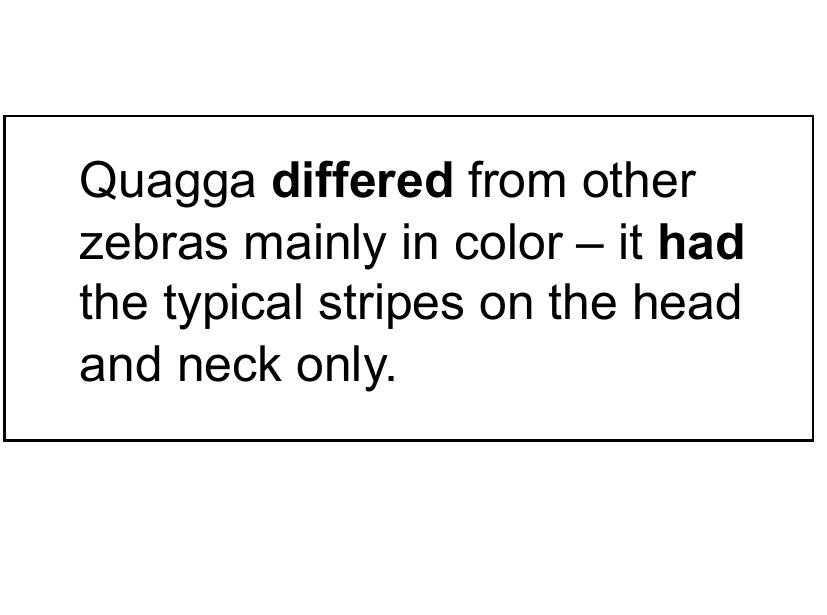}
	}
	\caption{Comparisons of protein and language. (a) Relationship between a MSA and the residue contact of one protein in the alignment. The positions that coevolved are highlighted in red and light blue. Residues within these positions where changes occurred are shown in blue. Given such a MSA, one can infer correlations statistically found between two residues that these sequence positions are spatially adjacent, i.e., they are contacts~\citep{Zerihun2018BiomolecularSP, ochoa2014practical, vorberg2017bayesian}. (b) One grammatically complex sentence contains long-distance dependencies (shown in bold).}
	\label{fig_03}
\end{figure}

Figure~\ref{fig_03_1} shows a MSA and its statistical inference for residue contact of one protein. This illustrates that long-range dependencies exist between two residues; they may be far apart in the sequence but close in space resulting in coevolution. Similarly, long-distance dependencies, which pose a problem for machine translation and RNNs, also appeared in languages. Figure~\ref{fig_03_2} shows an example of language grammar rules that require agreement in the category of words that might be far from each other~\citep{choshen2019automatically}. All these similarities illustrate that researchers can deal with protein data using successful methods in NLP.

However, proteins are not human languages, despite these and other similarities. For example, the training of LMs often requires a massively large corpus, which needs tokenization, i.e., splitting the text into individual tokens or directly using words as tokens, which serves computational goals and, ideally, can also fulfil linguistic goals in NLP~\citep{EthanCAlley2019UnifiedRP, AliMadani2021DeepNL, KevinKYang2018LearnedPE, EhsaneddinAsgari2015ContinuousDR, DanOfer2021TheLO}. Compared with algorithms in NLP, protein tokenization methods still remain at a low level without a well-defined and biologically meaningful protein token algorithm~\citep{MaiHaVu2022AdvancingPL}. This may be a direction for unlocking the secrets of proteins.
\section{Language Models}
This section firstly introduces encoder architectures of LMs broadly falling into two categories: recurrent neural networks (RNNs) and attention mechanisms, especially for long short-term memory (LSTM)~\citep{GuillaumeLample2016NeuralAF} and transformers~\citep{vaswani2017attention}. Then we present the commonly-used pre-trained LMs and their developments.
\subsection{Recurrent Neural Networks (RNNs) and Long Short-term Memory (LSTM)}
The first example of LM was studied by Andrey Markov, who proposed the Markov chain in 1913~\citep{hayes2013first,li2022language}. After that, some machine learning methods, hidden Markov models (HMMs) and their variants particularly, are described and applied as fundamental tools in many fields, including biological sequences ~\citep{bishop1986maximum} where the goal is to recover a data sequence that is not immediately observable~\citep{chiu2020scaling, gales2008application,nicolai2013solving,domingos2015master,stigler2011complex,wong2013dna}.

Neural networks started to produce superior results in various NLP tasks since 2010s~\citep{ferruz2022controllable}. RNNs allow previous outputs to be used as inputs while having hidden states to exhibit temporal dynamic behaviours. Therefore, RNNs can use their internal states to process variable-length sequences of inputs that are useful and applicable in NLP tasks~\citep{agmls2009novel}. 
\begin{figure}[t]
	\centering
	\subfloat[RNNs]
	{
		\label{fig4-a}
		\includegraphics[width=0.52\linewidth]{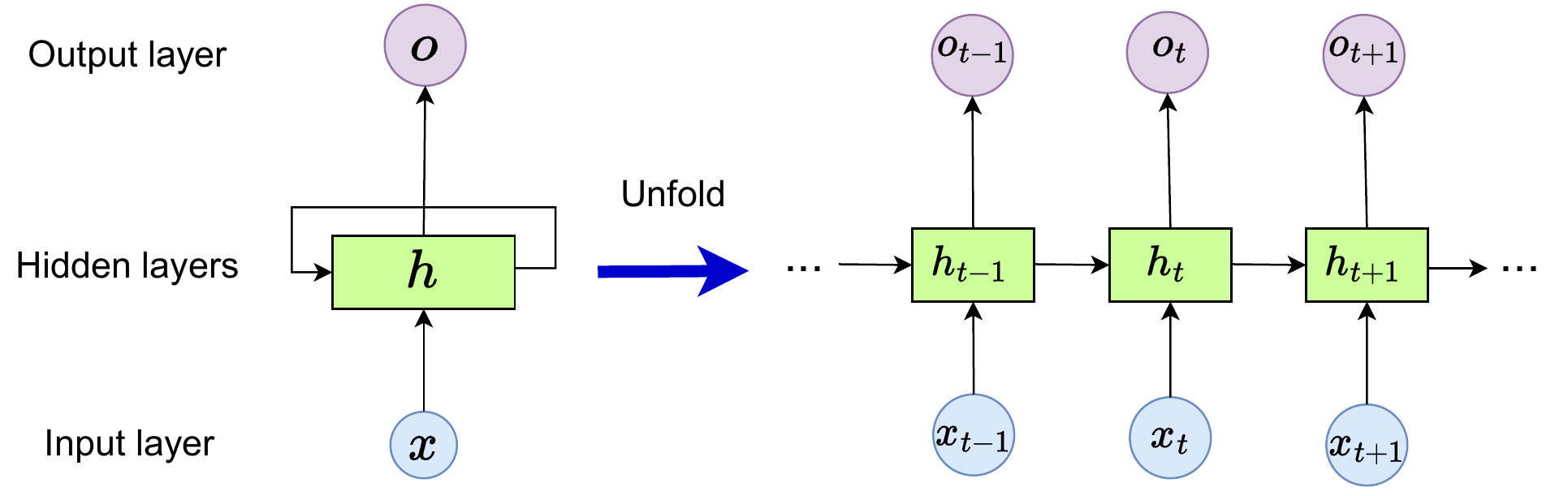}
	}
	\subfloat[LSTM cell]
	{
		\label{fig4-b}
		\includegraphics[width=0.42\linewidth]{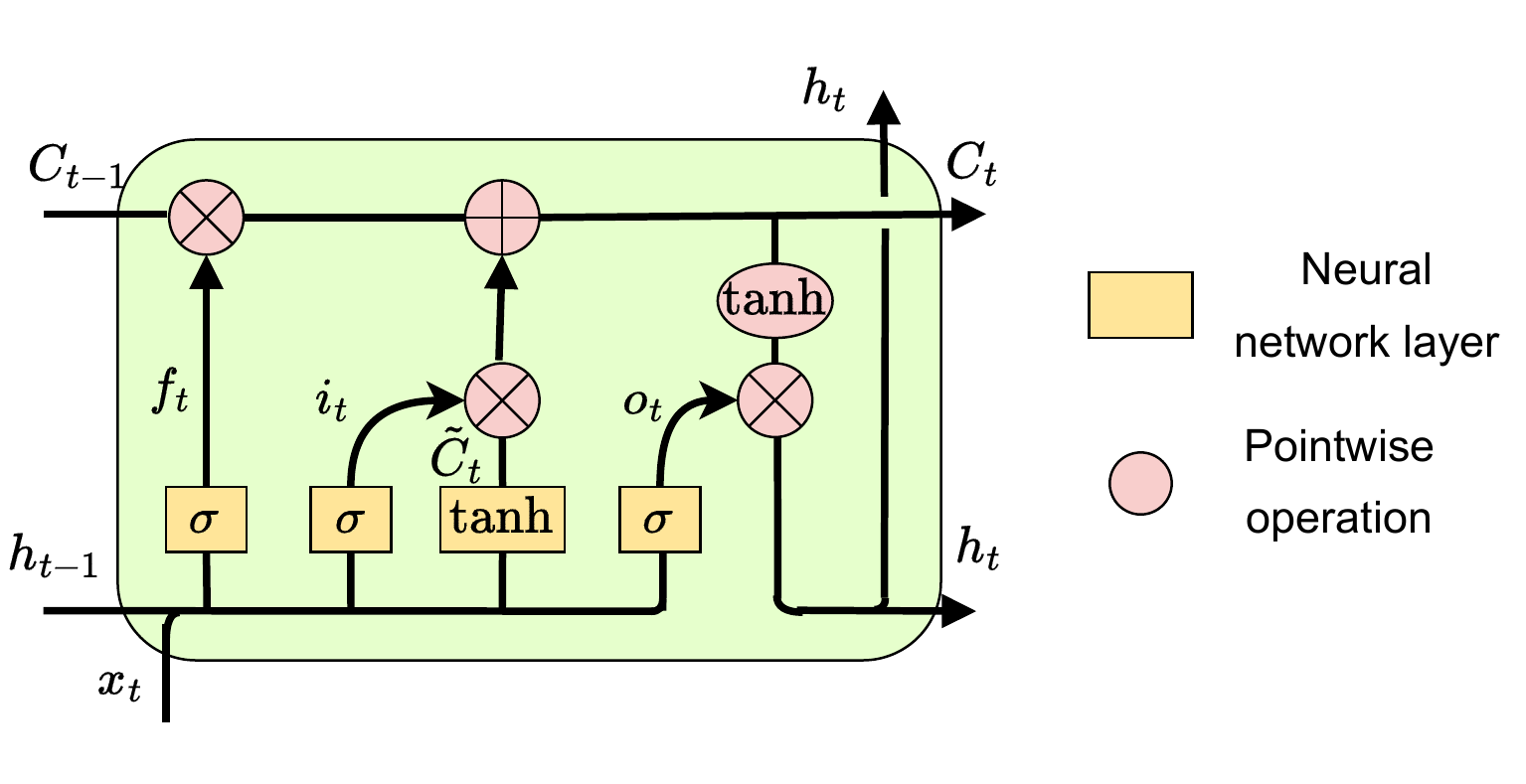}
	}
	\caption{Graphical explanation of RNNs and LSTM.}
	\label{fig_04}
\end{figure}

RNNs are typically shown in Figure~\ref{fig4-a}. In each timestep $t$, the input $x_{t}\in \mathbb{R}^l$, hidden $h_{t}\in \mathbb{R}^d$ and output state vectors $o_{t}\in \mathbb{R}^d$, where the superscripts $l$ and $d$ refer to the number of input features and the number of hidden units, respectively, are formulated as follows:
\begin{align*}
h_{t}&=g\left(W_{x}x_{t} + W_{h}h_{t-1} + b_h \right)\\
o_{t}&=g\left(W_{y}h_{t} + b_y \right)
\end{align*}
Where $W_x\in \mathbb{R}^{d\times l}$, $W_h\in \mathbb{R}^{d\times d}$ and $W_y\in \mathbb{R}^{d\times d}$ are the weights associated with the input, hidden and output vectors in the recurrent layer, and $b_h\in \mathbb{R}^{d}$, $b_y\in \mathbb{R}^{d}$ are the bias, which are shared temporally, $g$ is the activation function.

In order to deal with the vanishing gradient problem~\citep{hochreiter1991untersuchungen} that can be encountered when training traditional RNNs, LSTM networks were developed to process sequences of data. They presented superior capabilities in learning long-term dependencies~\citep{GuillaumeLample2016NeuralAF} with various applications such as time series prediction~\citep{schmidhuber2005evolino}, protein homology detection~\citep{hochreiter2007fast}, drug design~\citep{gupta2018generative}, etc. Unlike standard LSTM, bidirectional LSTM (BiLSTM) adds one more LSTM layer, reversing the information flow direction. This means it is capable of utilizing information from both sides and is also a powerful tool for modelling the sequential dependencies between words and phrases in a sequence~\citep{ma2021short}.

The LSTM architecture aims to provide a short-term memory that can last more timesteps, shown in Figure~\ref{fig4-b}, $\sigma$ and $\mathrm{tanh}$ represent the sigmoid and tanh layer. Forget gate layer in the LSTM is to decide what information is going to be thrown away from the cell state at timestep $t$, $x_{t}\in \mathbb{R}^l$, $h_t \in(-1,1)^d$ and $f_t \in(0,1)^d $ are the input, hidden state vectors and forget gate's activation vector. 
\begin{equation*}
f_t=\sigma(W_fx_t + U_fh_{t-1} +b_f )
\end{equation*}
Then, the input gate layer decides which values should be updated, and a tanh layer creates a vector of new candidate values, $\tilde{C}_t\in(-1,1)^d$ that could be added to the state, $i_t \in(0,1)^d$ is the input gate's activation vector.
\begin{align*}
i_t &=\sigma(W_ix_t + U_ih_{t-1} +b_i) \\
\tilde{C}_t &=\tanh (W_cx_t + U_ch_{t-1} +b_c)
\end{align*}
Next, we combine old state $C_{t-1}\in \mathbb{R}^d$ and new candidate values $\tilde{C}_t\in(-1,1)^d$ to create an update to the new state $C_t\in \mathbb{R}^d$.
\begin{equation*}
C_t=f_t \odot C_{t-1}+i_t \odot \tilde{C}_t
\end{equation*}
Finally, the output gate layer decides what parts of the cell state to be outputted, $o_t \in(0,1)^d$.
\begin{align*}
o_t &=\sigma (W_ox_t + U_oh_{t-1}+b_o) \\
h_t &=o_t \odot \tanh \left(C_t\right)
\end{align*}
where $\{W_f, W_i, W_c, W_o\} \in \mathbb{R}^{d \times l}, \{U_f, U_i, U_c, U_o\} \in \mathbb{R}^{d \times d}$ and $ \{b_f, b_i, b_c, b_o\} \in \mathbb{R}^d$ are weight matrices and bias vector parameters in the LSTM cell, $\odot$ means the pointwise multiplication.

\subsection{Attention Mechanism and Transformer}
Traditional Sequence-to-Sequence (Seq2Seq) models use RNNs or LSTMs as encoders and decoders~\citep{AlecRadford2017LearningTG} to process sequence and extract features for tasks. The final state of RNNs or LSTMs (better to say, encoders) must hold information for the entire input sequence, which may cause information loss. Therefore, traditional RNNs and LSTMs were soon superseded by attention mechanisms~\citep{DzmitryBahdanau2014NeuralMT, XuHan2021PreTrainedMP}, which help look at all hidden states from the encoder sequence for making predictions and were first applied for sequence modelling in machine translation~\citep{bahdanau2014neural}. 

The attention layer can access all previous states and learn their importance to weight them. Based solely on attention mechanisms, Google Brain released Transformer~\citep{vaswani2017attention} in 2017, a new network architecture dispensing with recurrence and convolutions entirely. This led to the development of pre-trained models, e.g., Bidirectional encoder representations from transformers (BERT)~\citep{JacobDevlin2022BERTPO} and Generative pre-trained transformer (GPT)~\citep{AlecRadford2022ImprovingLU,AlecRadford2022LanguageMA,TomBBrown2020LanguageMA}, which were trained with large language datasets. Unlike RNNs, Transformer processes the entire input all at once, using stacked self-attention layers for both the encoder and decoder. Each layer consists of a multi-head attention module followed by a feed-forward module with a residual connection and normalization. The vanilla single-head attention is called "Scaled Dot-Product Attention" and operates as follows:
\begin{equation*}
\operatorname{Att}(Q, K, V)=\operatorname{Softmax}\left(\frac{Q K^{\top}}{\sqrt{d}}\right) V
\end{equation*}
where $Q, K, V \in \mathbb{R}^{l \times d}$ are $d$-dimensional vector representations of $l$ words in sequences of queries, keys
and values, respectively. Multi-head attention allows the model to attend to information from different representation subspaces in parallel jointly. \begin{align*}
\operatorname{MultiHead}\left(Q, K, V\right)&=\mathrm{Concat}\left(\mathrm{head_1}, \ldots, \mathrm{head}_{\mathrm{h}}\right)W^O\\ 
\mathrm{where} \ \mathrm{head}_{\mathrm{i}}&=\mathrm{Att}\left(Q W_i^Q, K W_i^K, V W_i^V\right) 
\end{align*}
where the projections are parameter matrices $W_i^Q \in \mathbb{R}^{d \times d_i}, W_i^K \in \mathbb{R}^{d \times d_i}, W_i^V \in \mathbb{R}^{d \times d_i}$ and $W^O \in \mathbb{R}^{hd_i \times d}, d_i=d/h$, there are $h$ parallel attention layers or heads. Moreover, in Transformer, the position-wise feed-forward networks consist of two linear transformations with a ReLU activation in between, and positional encoding is added to the input embedding at the bottoms of the encoder and decoder stacks to make use of the order of the sequence.
\subsection{Pre-trained Language Models}
In order to train effective deep neural models focusing on storing knowledge for specific tasks with limited human-annotated data, transfer learning with a pre-training phrase and a fine-tuning stage has been adopted~\citep{XuHan2021PreTrainedMP,SebastianThrun1998LearningTL,SinnoJialinPan2010ASO}. In recent years, we have witnessed a rapid development of pre-trained LMs that have been widely used in NLP and computer vision, etc. Due to its prominent nature, the transformer gradually becomes a standard neural architecture for natural language understanding and generation. It also serves as the most commonly-used backbone neural architecture for pre-trained models as they have achieved state-of-the-art results on almost all NLP tasks. This indeed subverted our current perception of the performance of deep learning models, thus drawing more attention. An overview of some typical pre-trained LMs in general domains is shown in Table~\ref{Table_LMs}.
\begin{table*}[t]
	\caption{Representative pre-trained LMs in general domains}
	\label{Table_LMs}
	\setlength{\tabcolsep}{2.8pt}
	\centering
	\small
	\begin{adjustbox}{max width=\linewidth}
		\begin{threeparttable} 
			\begin{tabular}{lllll}
				\toprule 
				Model & Network & Objective &$\#$Params. &  Comments  \\
				\hline 
				ELMo~\citep{MatthewEPeters2018DeepCW}  &LSTM &Bidirectional LM & 93.6M & the first deep contextualized word representation\\
				GPT~\citep{AlecRadford2022ImprovingLU} &Transformer &Autoregressive LM & 110M & a pre-trained LM for predicting the next word\\
				BERT~\citep{JacobDevlin2022BERTPO}&Transformer& Masked LM &340M & the most commonly-used LM for predicting masked tokens\\
				Transformer-XL~\citep{ZihangDai2019TransformerXLAL} & Transformer  &Autoregressive LM & 257M & enabling learn dependencies beyond a fixed length \\
				XLM~\citep{GuillaumeLample2019CrosslingualLM} & Transformer& Multi-task &665M & cross-lingual pre-training \\
				Udify~\citep{DanKondratyuk201975L1} & BERT & Multi-task & $\sim$340M &leveraging a multilingual BERT self-attention model\\
				GPT-2~\citep{Radford2019} & Transformer& Autoregressive LM & 1.5B & larger model, more training data\\
				Grover~\citep{zellers2019grover}&GPT2 &Autoregressive LM & 1.5B 
				& defending against the general neural fake news\\
				XLNet~\citep{ZhilinYang2019XLNetGA} & BERT &Autoregressive LM & $\sim$340M & more training data, integrates ideas from Transformer-XL\\
				RoBERTa~\citep{YinhanLiu2019RoBERTaAR} & BERT & Masked LM & 355M & more training data, dynamic masking\\
				CTRL~\citep{NitishShirishKeskar2019CTRLAC} & Transformer &Autoregressive LM&1.63B& trained to control particular aspects of the generated text\\
				Megatron-LM~\citep{MohammadShoeybi2019MegatronLMTM} & Transformer & Autoregressive LM & 8.3B & a large transformer model\\
				ALBERT~\citep{ZhenzhongLan2019ALBERTAL} & BERT & Masked LM & 223M & a lite BERT\\
				DistillBERT~\citep{VictorSanh2019DistilBERTAD} & BERT & Masked LM & 65M & a distilled version of BERT\\
				SpanBERT~\citep{joshi2019spanbert} & BERT & Masked LM & $\sim$340M  & presenting and predicting the masked span \\
				MASS~\citep{KaitaoSong2019MASSMS} & Transformer & Seq2Seq LM & $\sim$307M & masked Seq2Seq pre-training\\
				MT-DNN~\citep{liu2019mt-dnn,liu2020mtmtdnn} & BERT & Multi-task & $\sim$340M &  for multiple natural language understanding tasks\\
				$\mathrm{MT-DNN}_{\mathrm{KD}}$~\citep{liu2019mt-dnn-kd} & MT-DNN & Multi-task & $\sim$340M & incorporating knowledge distillation \\
				ERNIE~\citep{ZhengyanZhang2019ERNIEEL} & BERT & Masked LM & $\sim$114M & incorporating knowledge graphs\\
				KnowBERT~\citep{MatthewEPeters2019KnowledgeEC} & BERT & Masked LM & $\sim$110M & incorporating knowledge bases into BERT \\
				KEPLER~\citep{XiaozhiWang2019KEPLERAU} & RoBERTa & Masked LM & $\sim$125M & incorporating knowledge embedding \\
				VideoBERT~\citep{ChenSun2019VideoBERTAJ} & BERT & Multimodal model & $\sim$340M & modelling between the visual and linguistic domain\\
				VisualBERT~\citep{LiunianHaroldLi2019VisualBERTAS} & BERT & Multimodal model & $\sim$110M & modelling a broad range of vision and language tasks\\
				ERNIE~\citep{YuSun2019ERNIEER} & BERT & Masked LM & $\sim$110M & a new masking strategy\\
				PEGASUS~\citep{zhang2020pegasus} & Transformer & Masked $\&$ Seq2Seq LM & 568M & for abstractive text summarization \\
				Unicoder-VL~\citep{GenLi2020UnicoderVLAU} & BERT & Multimodal model & $\sim$110M & cross-modal learning \\
				UNILM~\citep{unilmv2} & BERT & Bidirectional  $\&$ Seq2Seq LM & $\sim$110M & for natural language understanding and generation tasks\\
				Turing-NLG~\citep{JeffRasley2020DeepSpeedSO} & Transformer  &Autoregressive LM &17B& a hugely large LM\\
				ELECTRA~\citep{KevinClark2020ELECTRAPT} & BERT & Generator $\&$ Discriminator &335M &token detection\\
				GPT-3~\citep{TomBBrown2020LanguageMA} & GPT-2 & Autoregressive LM &175B & extending the model size\\
				T5~\citep{raffel2020exploring} & Transformer & Seq2Seq LM & 11B & producing new text as output\\
				Switch Transformer~\citep{WilliamFedus2021SwitchTS} & Transformer & Masked LM & 1.6T & increasing the pre-training speed \\
				BEIT~\citep{beit} & Transformer & Masked image model & 307M & a vision Transformer\\
				MT-NLG~\citep{ShadenSmith2022UsingDA} & Transformer & Autoregressive LM & 530B & the largest publicly monolithic transformer \\
				
				\toprule
			\end{tabular}
			\begin{tablenotes} 
				\item All examples report the largest model of their public series. Network displays high-level backbone models preferentially if they are used to initialize parameters. $\#$Param. means the number of parameters; M, millions; B, billions; T, trillions; $\&$ , and; $\sim$ means estimated data. Related terminologies are listed in Section~\ref{Notions and Terms}.
			\end{tablenotes} 
		\end{threeparttable}
	\end{adjustbox}
\end{table*}

\subsubsection{GPT and BERT}
With transformers as architectures and LM learning as objectives, BERT and GPT are the two landmarks that completely open the door towards the era of large-scale, deeply pre-trained LMs. 

GPT optimizes standard autoregressive language modelling during pre-training, which uses a transformer to model the conditional probability of each word and therefore, is powerful at predicting the next token in a sequence.
Formally, given an unsupervised corpus of tokens $\mathcal{X}=\left\{x_0, x_1, \ldots, x_n, x_n+1\right\}$, GPT applies a standard language modelling objective to maximize the following likelihood:
\begin{equation*}
\mathcal{L}(\mathcal{X})=\sum_{i=1}^{n+1} \log P\left(x_i \mid x_{i-k}, \ldots, x_{i-1} ; \Theta\right)
\end{equation*}
Where $k$ is the size of the context window, and the conditional probability $P$ is modelled using a network decoder with parameters $\Theta$. 

BERT uses a multi-layer bidirectional transformer encoder as its architecture. In the pre-training phase, BERT adopts the strategies of next-sentence prediction to understand sentence relationships with the help of a binary classifier and masked language modelling (MLM), which is powerful and applied in most self-supervised pre-training tasks. Formally, given a corpus consisting of tokens $\mathcal{X}=\left\{x_0, x_1, \ldots, x_n, x_n+1\right\}$, BERT maximizes the following likelihood: 
\begin{equation*}
\mathcal{L}(\mathcal{X})=\textstyle \sum_{x\in mask(\mathcal{X})} \log P ( x \mid \tilde{\mathcal{X}} ; \Theta)
\end{equation*}
where $mask(\mathcal{X})$ are the masked tokens, $\tilde{\mathcal{X}}$ is the result after masking some tokens in $\mathcal{X}$, and the probability $P$ is modeled by the transformer encoder with parameters $\Theta$.

\subsubsection{Post GPT and BERT Era}
After GPT and BERT, various of their improvements and variants have been proposed, as shown in Table~\ref{Table_LMs}. For example, researchers increased the size of models and datasets~\citep{YinhanLiu2019RoBERTaAR, ZhilinYang2019XLNetGA}, as large transformer models became the de facto standard in NLP on the basis of scaling laws, which can govern the dependence of overfitting on the model and dataset size for a given compute budget~\citep{JaredKaplan2020ScalingLF, JordanHoffmann2022TrainingCL}. The new masking strategies were proposed like entity-level masking, phrase-level masking~\citep{YuSun2019ERNIEER}, and span masking that masks the tokens consecutively according to the span length~\citep{joshi2019spanbert}, entity-level masking and phrase-level masking. Incorporating different data sources has also been developing as an important direction, such as utilizing multilingual corpora, and knowledge graphs~\citep{GuillaumeLample2019CrosslingualLM, MatthewEPeters2019KnowledgeEC}. Furthermore, because Pre-trained LMs are not data-hungry to labelled data and present better performance, they gradually stepped into different domains, including financial, computer vision and biomedical applications~\citep{araci2019finbert, BenyouWang2022PretrainedLM,YoshuaBengio2022ANP,TomasMikolov2013EfficientEO, lee2020biobert}, etc. 

\section{Protein Language Models}
As stated before, protein sequences corresponding to strings of amino-acid letters are a natural fit to most LMs, which are able to capture complex dependencies among these amino acids~\citep{OFER20211750}. pLMs have been developed and emerged as promising approaches for learning protein sequences. In this section, we first introduce LSTM pLMs, then present transformer pLMs with different implementation strategies and applications elaborately, especially the pLMs aimed at PSP. We listed a group of representative pLMs in Table~\ref{Table_pLMs}. The (pre-) training databases that appeared in this table are listed in Table~\ref{datasets}, where the CullPDB~\citep{GuoliWang2005PISCESRI} is a secondary structure prediction dataset.
\subsection{LSTM Protein Language Models}
\cite{MichaelSchantzKlausen2018NetSurfP20IP} trained a combination of convolutional and LSTM neural networks to predict protein structural features, solvent accessibility (SA), secondary structure (SS), structural disorder, and torsion angles ($\varphi$, $\psi$) for each residue of the input sequences. SPIDER3-Single~\citep{RhysHeffernan2018SinglesequencebasedPO} modelled on the single sequence instead of relying on evolutionary information from MSAs. Having similar training objectives and backbone architectures, such kind of examples are DeepPrime2Sec~\citep{EhsaneddinAsgari2019DeepPrime2SecDL}, SPOT-1D-Single~\citep{JaspreetSingh2021SPOT1DSingleIT}. Furthermore, DeepBLAST~\citep{JamesTMorton2020ProteinSA}, SPOT-1D-LM~\citep{JaspreetSingh2021SPOT1DLMRA} and SPOT-Contact-Single~\citep{JaspreetSingh2021SPOTContactSingleIS} are the usages of pre-trained pLMs to get embeddings for downstream tasks like contact map prediction, function prediction, etc.

However, \cite{RoshanRao2019EvaluatingPT} (TAPE) benchmarked a group of protein models across a panel of tasks, concluding that there exist opportunities for specific innovative design of protein models and training methods for vanilla LSTMs and Transformers. Methods specified for protein data processing have been researched and tested.

Without structural or evolutionary data, UniRep~\citep{EthanCAlley2019UnifiedRP} summarized arbitrary protein sequences into fixed-length vectors by multiplicative long-/short-term-memory (mLSTM)~\citep{BenKrause2016MultiplicativeLF}. Analogous to UniRep, UDSMProt~\citep{Strodthoff:2019universal} and SeqVec~\citep{MichaelHeinzinger2019ModelingTL} used LSTM or its variants to remember long-range dependencies for protein sequences to get rich representations that can be transferred and retrieved afterwards. ProSE~\citep{TristanBepler2021LearningTP} extra added structural supervision by residue-residue contact loss and structural similarity prediction loss to better capture the semantic organization of proteins and improved the ability to predict protein functions instead of using the masking loss only. Besides, \cite{TristanBepler2019LearningPS} proposed a soft symmetric alignment mechanism to measure similarities between sequence embeddings. CPCProt~\citep{AmyXLu2020SelfSupervisedCL} learned protein embeddings by formalizing the InfoNCE loss for the principle of mutual information maximization. All these models demonstrate the LSTM-like pLMs' ability to capture some biological properties of proteins.
\subsection{Transformer Protein Language Models}
\begin{figure}[t]
	\centering
	\subfloat[ESM-1b]
	{
		\label{ESM-1b}
		\includegraphics[width=0.95\linewidth]{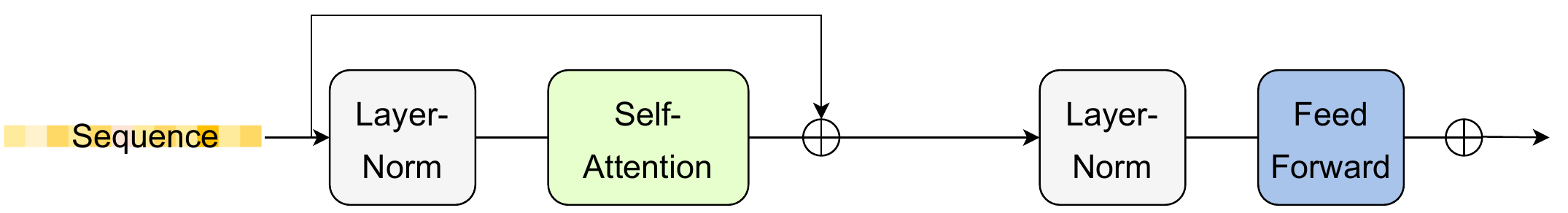}
	}\\
	\subfloat[ESM-MSA]
	{
		\label{ESM-MSA}
		\includegraphics[width=0.95\linewidth]{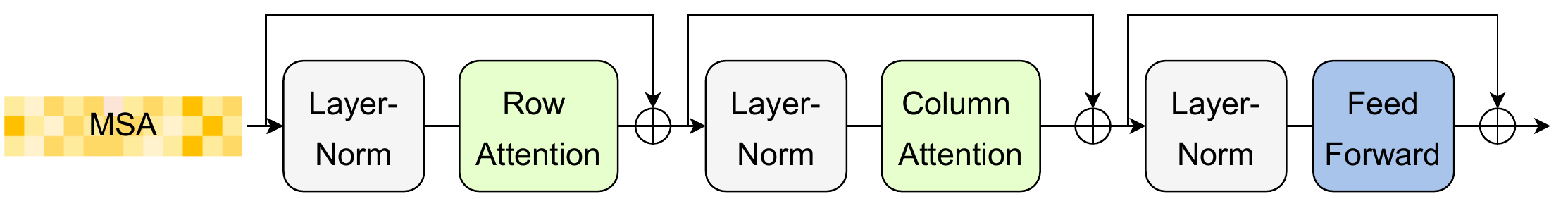}
	}
	\caption{Core modules of ESM-1b and ESM-MSA.}
	\label{ESM}
\end{figure}
\cite{AhmedElnaggar2021ProtTransTC} have trained six successful LMs (T5, ELECTRA, ALBERT, XLNet, BERT, and Transformer-XL, listed in Tabel~\ref{Table_LMs}) on protein sequences containing 393 billion amino acids using many resources (5616 GPUs and one TPU Pod). ESM-1b~\citep{AlexanderRives2019BiologicalSA} employed a deep Transformer (shown in Figure~\ref{ESM-1b}) and a masking strategy to build up complex representations that incorporate context from across the sequence. The results of ProtTrans~\citep{AhmedElnaggar2021ProtTransTC} and ESM-1b implied that large-scale pLMs have the advantage of learning the grammar of proteins even without using evolutionary information. Like approaches in NLP that change masking strategy, \cite{MatthewBAMcDermott2021AdversarialCP} updated the random masking scheme with a fully differentiable adversarial masking model. PMLM~\citep{LiangHe2022PretrainingCP} considered the dependency among masked tokens to capture the correlations (coevolution) of inter-residues, which was demonstrated to improve the performance on the TAPE contact benchmark. 
\subsubsection{Multi-source Knowledge Enhancement of Protein Representations}
Extra information, including MSAs, functions, structures and biological priors, may enrich protein embeddings. In detail, firstly, MSA Transformer~\citep{RoshanRao2021MSAT} extended the transformer LMs to deal with sets of sequences as input by alternating attention over rows and columns, shown in Figure~\ref{ESM-MSA}. Its internal representations enable high-quality unsupervised structure learning with an order of magnitude fewer parameters than contemporaneous pLMs. Secondly, ProteinBERT~\citep{DanOfer2021ProteinBERTAU} was pre-trained on protein sequences and Gene Ontology (GO) annotations which can be encoded as a binary vector. The learned embeddings contain information from both sequence and GO annotation to predict diverse protein functions~\citep{MAshburner2000GeneOT}; likewise, OntoProtein~\citep{NingyuZhang2022OntoProteinPP} considered GO as a factual knowledge graph, which was used to enhance protein representations. Finally, \cite{SanaaMansoor2021TowardMG} encoded two types of protein information (sequence and structure) through joint training in a semi-supervised manner. \cite{TristanBepler2021LearningTP} have carried out multi-task with structural supervision, leading to an even better-organized embedding space. STEPS~\citep{CanChen2022StructureawarePS} tried to correlate the embeddings learned from sequence and structure by pseudo bi-level optimization. 

However,~\cite{WeijieLiu2019KBERTEL} indicated that not all external knowledge is beneficial for downstream tasks. Thanks to the construction of benchmarks which can give relatively fair results of different methods. PEER benchmarks~\citep{MinghaoXu2022PEERAC} were built for protein sequence understanding, including protein function and structure prediction, protein-protein interaction prediction, protein-ligand interaction prediction tasks, etc. Its results showed that selecting suitable auxiliary tasks can boost different models' performance. Therefore, it is necessary to inject external features and design algorithms carefully and adequately.
\subsubsection{Transformer Models Designed for Protein Structure Prediction}
\begin{figure*}[h]
	\begin{center}
		\includegraphics[width=0.95\linewidth]{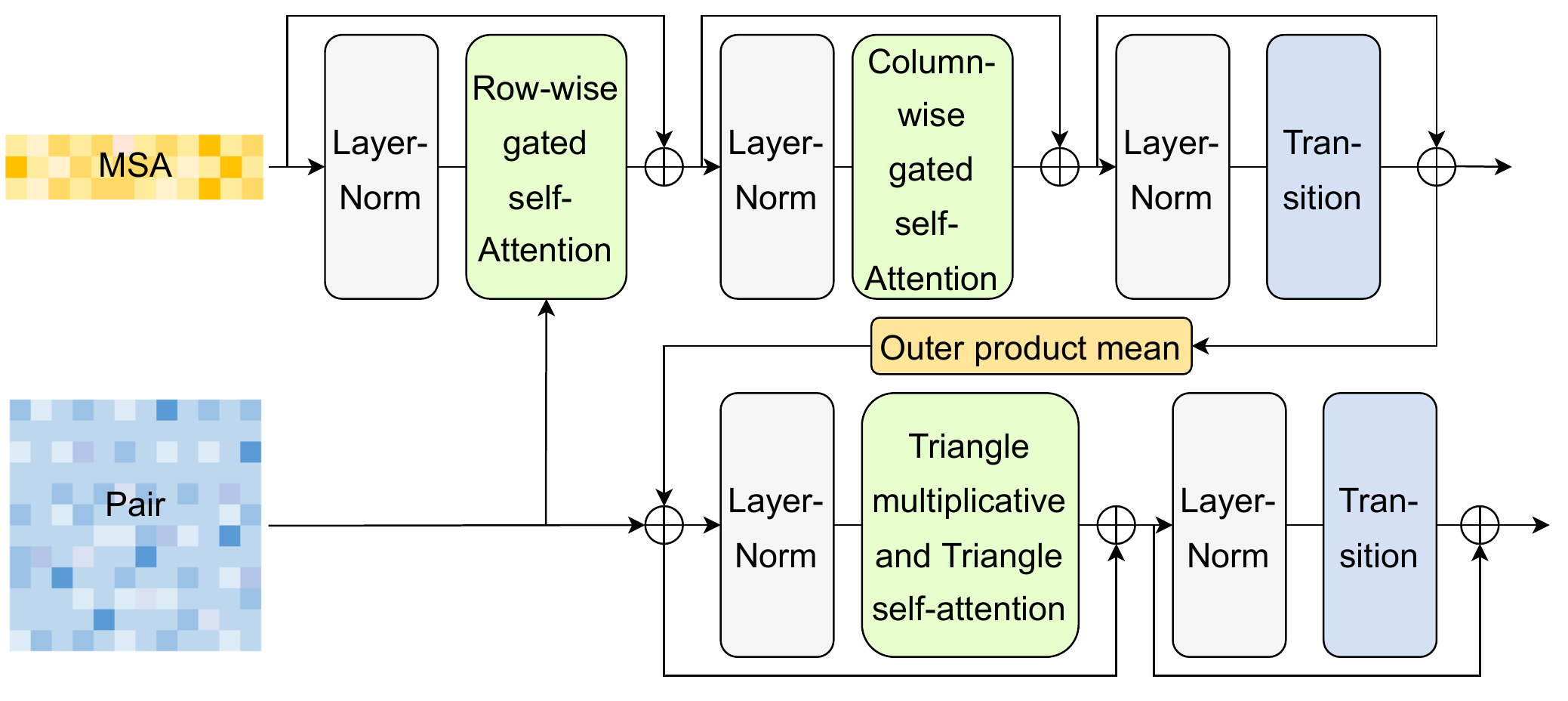}
	\end{center}
	\caption{Evoformer block.}
	\label{Evoformer}
\end{figure*}
Early pLMs tend to predict structural features~\citep{MichaelSchantzKlausen2018NetSurfP20IP, RhysHeffernan2018SinglesequencebasedPO, EhsaneddinAsgari2019DeepPrime2SecDL,JunwenLuo2020SelfSupervisedRL, JaspreetSingh2021SPOTContactSingleIS, RoshanRao2021TransformerPL, MagnusHaraldsonHie2022NetSurfP30AA, ShaunMKandathil2020DeepLP, MichaelHeinzinger2022ContrastiveLO}, like SS, SA, torsion angles, remote homology, and contact map, etc., which are useful for constructing protein 3D structures. Current pLMs tend to predict PSP end-to-end, which are introduced as follows. 

Evoformer is the core module (encoder) of the famous network, AF2~\citep{JohnMJumper2021HighlyAP}, which is repeated 48 times with no shared weights (shown in Figure~\ref{Evoformer}). It uses a variant of axial attention~\citep{JonathanHo2019AxialAI}, including row-wise gated self-attention with pair bias and column-wise gated self-attention, to process the MSA representation, which is transitioned by a 2-layer neural network, then used to update the pair representation by an outer product mean block, containing linear transforms, outer product, and mean, etc. In order to make the pair representation in the embedding space satisfy the demand of consistency, like the triangle inequality on distances, a triangle multiplicative update block and a triangle self-attention block were designed. The former updates the pair representation by combining information within each triangle of graph edges, while the latter operates self-attention around graph nodes. These ingenious designs let the output of Evoformer produce more insightful patterns for accurate PSP. Besides,~\cite{MingyangHu2022ExploringE} have shown that pLMs, especially those trained for PSP, like Evoformer, are valuable and general-purpose for various structure and function tasks. 

AminoBERT~\citep{RatulChowdhury2021SinglesequencePS} adopted a transformer to learn latent information from a single sequence, whose representation is inputted in a recurrent geometric network to generate the backbone structure of a protein. ESM-2~\citep{ZemingLin2022LanguageMO} is the largest pLM to date, extending ESM-1b and having parameters up to 15 billion, which has achieved the lowest validation perplexity and highest TM-score~\citep{YangZhang2005TMalignAP} on CASP14 compared with other smaller ESM models. The results indicated that improving the model sizes of pLMs is able to improve the PSP performance with little or no evolutionary information. OmegaPLM~\citep{RuidongWu2022HighresolutionDN} was trained with a stack of efficient GAU layers~\citep{WeizheHua2022TransformerQI} to get single- and pairwise-residue embeddings, which are expected to contain structural and functional information by different masking strategies, including random masking, sequential masking, and span masking~\citep{joshi2019spanbert}. In GAU, gate operation is applied after the attention aggregation and uses $\text{relu}^2(\cdot)$ to replace the conventional $\text{softmax}(\cdot)$, which performs better in terms of both computation speed and convergence rate in original experiments. AntiBERTy~\citep{JeffreyARuffolo2022FastAA} was pre-trained on antibody sequences to produce contextual embeddings for subsequent antibody structure prediction.

\subsubsection{Other Applications}
In terms of sequence generation, ProGen~\citep{madani2020progen} was trained on sequences conditioned on a set of protein properties like function or affiliation with a particular organism, and the training database contains about 280 million protein sequences and associated properties from different datasets. Compared with MSA Transformer~\citep{RoshanRao2021MSAT}, MSA2Prot~\citep{BeplerTristan2022FewSP} is a MSA-to-protein transformer, developed axial attention and cross attention for transformer encoder and decoder to model sequence probabilities autoregressively. Other methods include ProtGPT2~\citep{NoeliaFerruz2022ADU},  RITA~\citep{hesslow2022rita}, and ProGen2~\cite{ErikNijkamp2022ProGen2ET} (up to 6.4 billion parameters), etc.

Other than the models mentioned above, a group of pLMs are developed and trained for different purposes. For example, 
ESM-1v~\citep{meier2021language}, Tranception~\citep{PascalNotin2022TranceptionPF} enabled the prediction of the effects of mutations. Structured Transformer~\citep{JohnIngraham2019GenerativeMF}, ESM-IF1~\citep{hsu2022learning}, Fold2Seq~\citep{YueCao2021Fold2SeqAJ}, PeTriBERT~\citep{BaldwinDumortier2022PeTriBERTA}, ProteinMPNN~\citep{JDauparas2022RobustDL}, and~\citep{ZhangyangGao2022AlphaDesignAG, gao2022pifold, tan2022generative}, etc., were proposed for protein design, which is also referred to as the inverse protein folding problem, meaning recovering the native sequence of a protein from its tertiary structure (coordinates). ProtTucker~\citep{MichaelHeinzinger2022ContrastiveLO} utilized pLM and contrastive learning~\cite{chen2020simple,xia2022simgrace} strategy to improve the ability to recognize distant homologous, and other tasks, including profile prediction~\citep{PascalSturmfels2020ProfilePA}, evolutionary velocity prediction~\citep{BrianHie2021EvolutionaryVW}, enzymatic active sites identification~\citep{LocKwateDassi2021IdentificationOE}, etc. 

\begin{table*}[t]
	\caption{List of representative pLMs}
	\label{Table_pLMs}
	\setlength{\tabcolsep}{2.8pt}
	\centering
	\small
	\begin{adjustbox}{max width=\linewidth}
		\begin{threeparttable} 
			\begin{tabular}{llllllll}
				\toprule 
				Model and Repository&Approach & Input & Network &$\#$Embedding &$\#$Param.  & Pre-training Database \\
				\hline 
				\href{https://services.healthtech.dtu.dk/service.php?NetSurfP-2.0}{NetSurfP-2.0}~\citep{MichaelSchantzKlausen2018NetSurfP20IP} & Supervised & MSA, Structure & CNN, BiLSTM & 2048 & N/A & PDB, UniRef30 \\
				
				\href{http://sparks-lab.org}{SPIDER3-Single}~\citep{RhysHeffernan2018SinglesequencebasedPO} & Supervised & Sequence, Structure & LSTM-BRNN~\citep{RhysHeffernan2017CapturingNI} & 1024, 512 & N/A & 12442 proteins \\
				\href{https://github.com/Rostlab/SeqVec}{SeqVec}~\citep{MichaelHeinzinger2019ModelingTL}&Unsupervised  &Sequence & ELMo & 1024 &$\sim$93.6M & UniRef50\\
				
				\href{https://github.com/churchlab/Unirep}{UniRep}~\citep{EthanCAlley2019UnifiedRP}&Unsupervised & Sequence & mLSTM~\citep{BenKrause2016MultiplicativeLF} & 1900 &$\sim$18.2M & UniRef50\\
				
				\href{https://github.com/tbepler/protein-sequence-embedding-iclr2019}{SSA}~\citep{TristanBepler2019LearningPS}& Supervised & Sequence, Structure & BiLSTM & 100, 512 & N/A & Pfam, SCOP\\
				
				\href{https://github.com/ehsanasgari/DeepPrime2Sec}{DeepPrime2Sec}~\citep{EhsaneddinAsgari2019DeepPrime2SecDL} & Supervised & MSA, Sturcture & ELMo, CNN, BiLSTM & N/A &  N/A & UniRef50, Swiss-Prot, CullPDB \\
				
				\cite{JohnIngraham2019GenerativeMF} & Unsupervised & Structure & Transformer & 128 & N/A & CATH4.2 \\
				
				\multirow{2}{*}{\href{https://github.com/songlab-cal/tape}{TAPE}~\citep{RoshanRao2019EvaluatingPT}} & \multirow{2}{*}{Unsupervised} & \multirow{2}{*}{Sequence} & LSTM& 2048 & N/A & \multirow{2}{*}{Pfam} \\
				&&&Transformer & 768& 38M & \\
				\href{https://github.com/facebookresearch/esm}{ESM-1b}~\citep{AlexanderRives2019BiologicalSA} & Unsupervised & Sequence & Transformer & 1280& 650M & UniParc\\
				
				\href{https://github.com/nstrodt/UDSMProt}{UDSMProt}~\citep{Strodthoff:2019universal} & Unsupervised & Sequence & LSTM & 400 & $\sim$24M & Swiss-Prot\\
				
				\href{https://github.com/amyxlu/CPCProt}{$\mathrm{CPCProt}_{\mathrm{GRU \_ large}}$}~\citep{AmyXLu2020SelfSupervisedCL} & Unsupervised & Sequence & GRU~\citep{cho2014properties} & 1024 & 8.4M & Pfam \\
				\href{https://github.com/amyxlu/CPCProt}{$\mathrm{CPCProt}_{\mathrm{LSTM}}$}~\citep{AmyXLu2020SelfSupervisedCL} & Unsupervised & Sequence & LSTM & 2048 & 71M & Pfam \\
				
				\cite{PascalSturmfels2020ProfilePA} & Supervised & MSA & Transformer & N/A & N/A & Pfam \\
				
				\href{https://github.com/lucidrains/progen}{ProGen}~\citep{madani2020progen} & Unsupervised & Sequence, Property & Transformer & 1028 & 1.2B & $\sim$280M proteins\\
				
				
				\href{https://github.com/agemagician/ProtTrans}{ProtTXL}~\citep{AhmedElnaggar2021ProtTransTC} & Unsupervised & Sequence & Transformer-XL & 1024 & 562M & BFD100, UniRef100 \\
				\href{https://github.com/agemagician/ProtTrans}{ProtBert}~\citep{AhmedElnaggar2021ProtTransTC} & Unsupervised & Sequence & BERT & 1024 & 420M & BFD100, UniRef100 \\
				\href{https://github.com/agemagician/ProtTrans}{ProtXLNet}~\citep{AhmedElnaggar2021ProtTransTC} & Unsupervised & Sequence & XLNet & 1024 & 409M &UniRef100\\
				\href{https://github.com/agemagician/ProtTrans}{ProtAlbert}~\citep{AhmedElnaggar2021ProtTransTC} & Unsupervised & Sequence & ALBERT & 4096 & 224M &UniRef100\\
				\href{https://github.com/agemagician/ProtTrans}{ProtElectra}~\citep{AhmedElnaggar2021ProtTransTC} & Unsupervised & Sequence & ELECTRA & 1024 & 420M &UniRef100\\
				\href{https://github.com/agemagician/ProtTrans}{ProtT5}~\citep{AhmedElnaggar2021ProtTransTC} & Unsupervised & Sequence & T5 & 1024 & 11B &UniRef50, BFD100\\
				
				\href{https://github.com/jas-preet/SPOT-1D-Single}{SPOT-1D-Single}~\citep{JaspreetSingh2021SPOT1DSingleIT} & Supervised & Sequence, Structure & BiLSTM, ResNet~\citep{KaimingHe2016IdentityMI} & 256 & N/A & 39120 proteins \\
				
				\href{https://github.com/tbepler/prose}{ProSE}~\citep{TristanBepler2021LearningTP} &Supervised & Sequence, Structure & BiLSTM & 1024 & 1M & UniRef90, SCOPe \\
				
				\href{https://github.com/facebookresearch/esm}{MSA Transformer}~\citep{RoshanRao2021MSAT} & Unsupervised & MSA & Transformer & 768 & 100M & UniRef50, UniClust30\\
				\href{https://github.com/facebookresearch/esm}{ESM-1v}~\citep{meier2021language} & Unsupervised & Sequence & ESM-1b & 1280 & 650M & UniRef90\\
				\href{https://github.com/facebookresearch/esm}{ESM-IF1}~\citep{hsu2022learning} & Supervised & Sequence, Structure &GVP~\citep{BowenJing2020LearningFP}, Transformer & 512 & 142M & UniRef50, CATH\\
				\multirow{2}{*}{\href{https://github.com/nadavbra/protein_bert}{ProteinBERT}~\citep{DanOfer2021ProteinBERTAU}} & \multirow{2}{*}{Supervised} & Sequence &  \multirow{2}{*}{Transformer} & \multirow{2}{*}{128, 512} & \multirow{2}{*}{$\sim$16M} & \multirow{2}{*}{UniRef90} \\
				& & GO annotation & & & &\\
				
				\href{https://github.com/IBM/fold2seq}{Fold2Seq}~\citep{YueCao2021Fold2SeqAJ} & Supervised & Sequence, Structure & Transformer & 256 & N/A & CATH4.2 \\
				
				\href{https://github.com/aqlaboratory/rgn2}{AminoBERT}~\citep{RatulChowdhury2021SinglesequencePS} & Unsupervised & Sequence & Transformer & 3072 & N/A & UniParc \\
				\href{https://github.com/deepmind/alphafold}{Evoformer}~\citep{JohnMJumper2021HighlyAP} & Supervised & MSA, Structure & Attention network &  384, 128 & 93M & PDB, BFD, UniClust30, etc.\\
				\href{https://github.com/HeliXonProtein/OmegaFold}{OmegaPLM}~\citep{RuidongWu2022HighresolutionDN}& Unsupervised & Sequence & GAU~\citep{WeizheHua2022TransformerQI} & 1280 & 670M & UniRef50 \\
				\href{https://github.com/zjunlp/OntoProtein}{OntoProtein}~\citep{NingyuZhang2022OntoProteinPP} & Supervised & Sequence, GO & ProtBert, BERT & 1024 & N/A & ProteinKG25 \\
				PeTriBERT~\citep{BaldwinDumortier2022PeTriBERTA} & Unsupervised & Sequence, Structure & BERT & 3072 & <40M & AlphaFoldDB \\
				MSA2Prot~\citep{BeplerTristan2022FewSP} & Unsupervised & MSA & Transformer & 768 & N/A & Pfam \\
				
				PMLM~\citep{LiangHe2022PretrainingCP} & Unsupervised & Sequence & Transformer & 1280 & 715M & UniRef50 \\
				
				\href{https://github.com/salesforce/progen}{ProGen2}~\citep{ErikNijkamp2022ProGen2ET} & Unsupervised & Sequence & Transformer & 4096 & 6.4B & UniRef90, BFD30\\
				\href{https://github.com/OATML-Markslab/Tranception}{Tranception}~\citep{PascalNotin2022TranceptionPF} & Unsupervised & Sequence & Transformer & 1280 & 700M & UniRef100 \\
				
				\href{https://huggingface.co/nferruz/ProtGPT2}{ProtGPT2}~\citep{NoeliaFerruz2022ADU} & Unsupervised & Sequence & GPT-2 & 1280 & 738M&  UniRef50 \\
				\href{https://github.com/lightonai/RITA}{RITA}~\citep{hesslow2022rita} & Unsupervised & Sequence & GPT-3 & 2048 & 1.2B & UniRef100 \\
				\href{https://github.com/facebookresearch/esm}{ESM-2}~\citep{ZemingLin2022LanguageMO} & Unsupervised & Sequence & Transformer & 5120 & 15B & UniRef50 \\

				
				\toprule
			\end{tabular}
			\begin{tablenotes} 
				\item All examples report the largest model of their public series, the model name with colour is linked with GitHub or server page. Approach and database are listed for the pre-training stage, and the latter is elaborated on in Section~\ref{Databases}. Input is classified into protein sequence, MSA, structure (structural features or coordinates), and function. Network displays high-level backbone models preferentially if they are used. $\#$Embedding means the dimension of embeddings; $\#$Param., the number of parameters of network; M, millions; B, billions; T, trillions; N/A, null; $\&$, and; $<$, less than; $\sim$ estimated data. 
			\end{tablenotes} 
		\end{threeparttable}
	\end{adjustbox}
\end{table*}

\section{Methods of Protein Structure Prediction (PSP)}
This section mainly introduces different methods that work for different levels of prediction of protein structures, including traditional methods and deep learning methods. Among them, pLM-based models function importantly and significantly influence protein tasks.

\tikzstyle{leaf}=[draw=hiddendraw,
rounded corners,minimum height=1.2em,
fill=hidden-orange!40,text opacity=1, align=center,
fill opacity=.5,  text=black,align=left,font=\scriptsize,
inner xsep=3pt,
inner ysep=1pt,
]
\begin{figure*}[t]
	\centering
	\begin{forest}
		for tree={
			forked edges,
			grow=east,
			reversed=true,
			anchor=base west,
			parent anchor=east,
			child anchor=west,
			base=middle,
			font=\footnotesize,
			rectangle,
			draw=hiddendraw,
			rounded corners,align=left,
			minimum width=2.5em,
			minimum height=1.2em,
			s sep=6pt,
			inner xsep=3pt,
			inner ysep=1pt,
		},
		where level=1{text width=4.5em}{},
		where level=2{text width=6em,font=\scriptsize}{},
		where level=3{font=\scriptsize}{},
		where level=4{font=\scriptsize}{},
		where level=5{font=\scriptsize}{},
		[Structural\\Features, edge=hiddendraw
		[1D\\Features,text width=3em, edge=hiddendraw
		[ANN, text width=3.8em, edge=hiddendraw
		[
		\cite{LHHolley1989ProteinSS}{,}
		\cite{BurkhardRost1993ImprovedPO}{,}
		DBNN~\citep{XinQiuYao2008ADB}{,}\\
		DNSS~\citep{MatthewSpencer2015ADL}{,}
		\cite{HarisHasic2017AHM}{,}
		\href{http://sparks-lab.org/}{SPIDER2}~\citep{RhysHeffernan2015ImprovingPO}
		,leaf,text width=25.4em, edge=hiddendraw]
		]
		[CNN, text width=3.8em, edge=hiddendraw
		[
		\cite{AkosuaBusia2016ProteinSS}{,}
		MUST-CNN~\citep{ZemingLin2016MUSTCNNAM}{,}\\
		\href{http://raptorx2.uchicago.edu/StructurePropertyPred/predict/}{RaptorX-Property}~\citep{ShengWang2016RaptorXPropertyAW}{,}
		GSN~\citep{JianZhou2014DeepSA}
		,leaf,text width=25.4em, edge=hiddendraw]
		]
		[RNN, text width=3.8em, edge=hiddendraw
		[\href{http://scratch.proteomics.ics.uci.edu/}{SSpro/ACCpro5}~\citep{ChristopheMagnan2014SSproACCpro5A}{,}
		\href{http://sparkslab.org}{SPIDER3}~\citep{RhysHeffernan2017CapturingNI}{,}\\
		\cite{SrenKaaeSnderby2014ProteinSS}{,}
		\href{http://sparks-lab.org}{SPIDER3-Single}~\citep{RhysHeffernan2018SinglesequencebasedPO}{,}\\
		\href{http://distilldeep.ucd.ie/porter/}{Porter 5}~\citep{MirkoTorrisi2018Porter5S}
		,leaf,text width=25.4em, edge=hiddendraw]
		]
		[Hybrid, text width=3.8em, edge=hiddendraw
		[
		\href{https://services.healthtech.dtu.dk/service.php?NetSurfP-2.0}{NetSurfP-2.0}~\citep{MichaelSchantzKlausen2018NetSurfP20IP}{,}
		\href{https://github.com/ehsanasgari/DeepPrime2Sec}{DeepPrime2Sec}~\citep{EhsaneddinAsgari2019DeepPrime2SecDL}{,}\\
		MASSP~\citep{BianLi2020AMD}{,}
		\href{https://sparks-lab.org/server/spot-1d-single/}{SPOT-1D-Single}~\citep{JaspreetSingh2021SPOT1DSingleIT}{,}
		\href{http://qianglab.scst.suda.edu.cn/crrnn2/}{CRRNN2}~\citep{BuzhongZhang2021MultitaskDL}
		,leaf,text width=25.4em, edge=hiddendraw]
		]
		[Transformer, text width=3.8em, edge=hiddendraw
		[
		SPOT-1D-LM~\citep{JaspreetSingh2021SPOT1DLMRA}{,}
		,leaf,text width=25.4em, edge=hiddendraw]
		]
		]
		[2D\\Features,text width=3em, edge=hiddendraw
		[ANN, text width=3.8em, edge=hiddendraw
		[
		\href{https://predictioncenter.org/}{MetaPSICOV2}~\citep{DanielWABuchan2018ImprovedPC}{,}
		DeepCDpred~\citep{ShuangxiJi2018DeepCDpredID}
		,leaf,text width=25.4em, edge=hiddendraw]
		]
		[CNN, text width=3.8em, edge=hiddendraw
		[
		RaptorX-Contact~\citep{ShengWang2016AccurateDN}{,}
		\href{https://github.com/multicom-toolbox/DNCON2/}{DNCON2}~\citep{BadriAdhikari2017DNCON2IP}{,}\\
		\href{https://github.com/largelymfs/deepcontact}{DeepContact}~\citep{YangLiu2017EnhancingEC}{,}
		\href{https://github.com/psipred/DeepCov}{DeepCov}~\citep{DavidTJones2018HighPI}{,}
		\cite{JinLi2021StudyOR}{,}\\
		\href{https://github.com/multicom-toolbox/deepdist}{DeepDist}~\citep{TianqiWu2020DeepDistRI}{,}
		\href{https://github.com/ba-lab/pdnet/}{PDNET}~\citep{BadriAdhikari2020AFO}{,}
		\href{https://github.com/multicom-toolbox/deepdist}{REALDIST}~\citep{BadriAdhikari2020REALDISTRP}{,}\\
		\href{https://seq2fun.dcmb.med.umich.edu//TripletRes/}{TripletRes}~\citep{YangLi2020DeducingHP}{,}
		\href{https://github.com/ElofssonLab/PconsC4}{PconsC4}~\citep{MircoMichel2019PconsC4FA}{,}
		\href{http://huanglab.phys.hust.edu.cn/DeepHomo/}{DeepHomo}~\citep{YumengYan2021AccuratePO}
		,leaf,text width=25.4em, edge=hiddendraw]
		]
		[Hybrid, text width=3.8em, edge=hiddendraw
		[
		\href{http://sparks-lab.org/jack/server/SPOT-Contact/}{SPOT-Contact}~\citep{JackHanson2018AccuratePO}{,}
		\cite{ChenChen2020CombinationOD}
		,leaf,text width=25.4em, edge=hiddendraw]
		]
		[Transformer, text width=3.8em, edge=hiddendraw
		[\href{https://github.com/ jas-preet/SPOT-Contact-Single}{SPOT-Contact-Single}~\citep{JaspreetSingh2021SPOTContactSingleIS}{,}
		\cite{NicholasBhattacharya2021SingleLO}
		,leaf,text width=25.4em, edge=hiddendraw]
		]
		]	
		[Both,text width=3em, edge=hiddendraw
		[CNN, text width=3.8em, edge=hiddendraw
		[
		\href{https://github.com/psipred/DMPfold}{DMPfold}~\citep{JoeGGreener2018DeepLE}{,}
		\href{https://github.com/ShenLab/ThreaderAI}{ThreaderAI}~\citep{HaicangZhang2020TemplatebasedPO}
		,leaf,text width=25.4em, edge=hiddendraw]
		]
		[Hybrid, text width=3.8em, edge=hiddendraw
		[
		DMPfold2~\citep{ShaunMKandathil2020DeepLP}
		,leaf,text width=25.4em, edge=hiddendraw]
		]
		]	
		]
	\end{forest}
	\caption{Taxonomy of structural feature prediction methods. ANN, Artificial Neural Network; Hybrid means models use CNN- and RNN-based methods simultaneously. The model name with colour is linked with the official GitHub or server page.}
	\label{taxonomy_of_structural_methods}
\end{figure*}
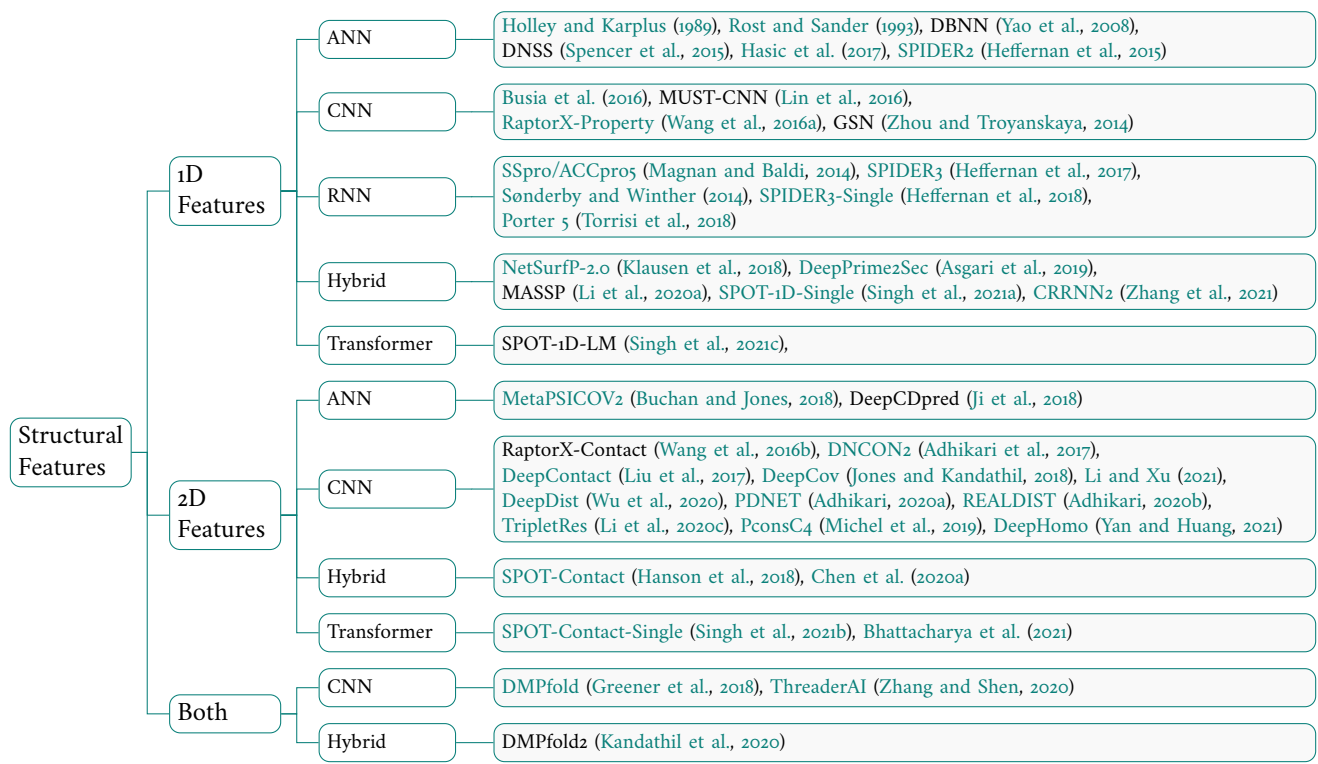

\subsection{Structural Features Prediction}
Structural features include 1D features (SS, SA, torsion angles, contact density, etc.) and 2D features (contact map and distance map)~\citep{MirkoTorrisi2020DeepLM}, which are useful for predicting protein structures. In the early stages of PSP, instead of predicting atom coordinates directly, researchers tried to predict structural features to evaluate their methods and aid the process of PSP. Methods of structural feature prediction are classified in Figure~\ref{taxonomy_of_structural_methods}. 

In the late 1980s, \cite{LHHolley1989ProteinSS} have already worked for protein secondary structure prediction (SSP) based on a neural network. DBNN~\citep{XinQiuYao2008ADB} combined dynamic Bayesian networks and neural networks to achieve improvements in SSP. DeepCNF~\citep{ShengWang2015ProteinSS} integrated conditional random fields (CRF) and shallow convolutional neural networks to predict SS of 3-state (H, E, C) and 8-state (G, H, I, E, B, T, S, L). \cite{RhysHeffernan2017CapturingNI} employed LSTM and bidirectional recurrent neural networks (BRNNs), which are capable of capturing long-range interactions. MASSP~\citep{BianLi2020AMD} used multi-tiered neural networks, which are composed of a CNN and a LSTM neural network, to get 1D structural attributes. As for 2D structural features prediction, RaptorX-Contact~\citep{ShengWang2016AccurateDN} integrated sequence and evolutionary coupling information by two deep residual neural networks to predict contact maps.~\cite{ChenChen2020CombinationOD} presented an attention-based convolutional neural network for protein contacts, including a sequence attention module and a regional attention network. \cite{JinLi2021StudyOR} studied the inter-atom distance and inter-residue orientation by a ResNet network and then built 3D structure models constrained by the predicted mean and deviation through PyRoseatta~\citep{SidharthaChaudhury2010PyRosettaAS}. Similarly, DeepDist~\citep{TianqiWu2020DeepDistRI}, PDNET~\citep{BadriAdhikari2020AFO}, REALDIST~\citep{BadriAdhikari2020REALDISTRP}, etc., were also proposed to learn the real‑value inter‑residue distance by designing regression models, which is more difficult than multi-class classification problem~\citep{JinboXu2018DistancebasedPF} but stepping forward for PSP.

However, the structural feature prediction tasks do not generally have significant practical values since there is already highly accurate 3D structure data from AF2. Is it a thing of the past? Considering it from another perspective, it can still provide a reference for the results of various proposed methods and work for finding the relationships between protein sequence, structure, and function by mining more protein grammars. For example, \cite{NicholasBhattacharya2021SingleLO} have introduced a simplified attention layer, factored attention, to find the role of attention, which achieved nearly identical performance to the Potts model with far fewer parameters. Particularly, DeepHomo~\citep{YumengYan2021AccuratePO} aimed to predict inter-protein residue-residue contacts across homo-oligomeric protein interfaces by integrating multi-source information to remove the potential intra-protein contact noises that exist in the MSA. Geoformer~\citep{RuidongWu2022HighresolutionDN} refined the details of contact prediction to illustrate its effectiveness in resolving the problem of triangular inconsistency.

\subsection{Traditional Methods for PSP}
\cite{KimTSimons1997AssemblyOP} generated native-like structures from fragments of unrelated protein structures using Bayesian scoring functions and a simulated annealing procedure in the 1990s. TOUCHSTONE II~\citep{YangZhang2003TOUCHSTONEIA} proposed a parallel hyperbolic sampling algorithm used in the Monte Carlo simulation processes to accelerate the conformational search faster. \cite{CarolARohl2004ProteinSP} utilized the Rosetta algorithm for \emph{de novo} PSP, which can assemble fragments by a Monte Carlo strategy. Such a strategy was also adopted in QUARK~\citep{DongXu2012AbIP}. \cite{AndreaCavalli2007ProteinSD} used chemical shifts as structural restraints to help determine the conformations of proteins. Based on backbone chemical shifts, \cite{ChristopheSchmitz2012ProteinSD} stepped further on the quest for reliable PSP using pseudocontact shifts. EdaFold~\citep{DavidSimoncini2012APF} is a fragment-based PSP method via estimation of distribution algorithm that is learned from previously generated decoys. Stepping on this, a cluster-based model and an energy-based variation were provided by \cite{DavidSimoncini2017BalancingEA}. UniCon3D~\citep{DebswapnaBhattacharya2016UniCon3DDN} is a generative, probabilistic model using united-residue conformational search, sampling lower energy conformations with higher accuracy than traditional random sampling. 

Structural features like SS and contact maps appear to help the PSP. DCAFold~\citep{JoannaISulkowska2012GenomicsaidedSP} integrated contacts estimated from direct coupling analysis with an accurate knowledge of local information to fold proteins. FragFold ~\citep{TomaszKosciolek2014DeNS} used fragment assembly with both statistical potentials and predicted contacts. RASREC~\citep{TatjanaBraun2015CombiningEI} integrated evolutionary information in the form of intra-protein residue-residue contacts. CONFOLD~\citep{BadriAdhikari2015CONFOLDRC} developed a distance geometry algorithm using structural features as restraints. \cite{SergeyOvchinnikov2016ImprovedDN} used structural information during Rosetta conformational sampling and refinement to improve the model's accuracy. Besides, RBO Aleph~\citep{MahmoudMabrouk2015RBOAL} leveraged evolutionary and physicochemical information to predict contacts, used in conformational space search,
afterwards, similar instances are SCDE~\citep{GuiJunZhang2020SecondarySA}, TDFO~\citep{GuiJunZhang2020TwoStageDF}. 

\subsection{Deep Learning Methods for PSP: Past, Present, and Future}
\subsubsection{Structural Features-Based Methods}
\begin{figure*}[h]
	\begin{center}
		\includegraphics[width=0.95\linewidth]{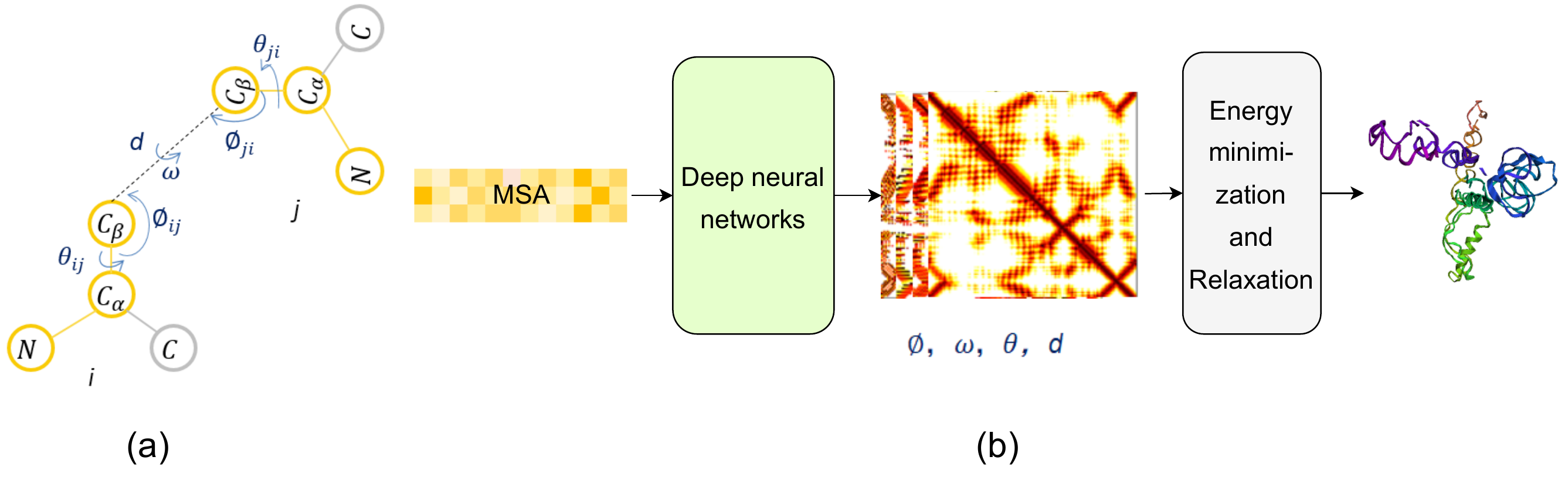}
	\end{center}
	\caption{A pipeline of predicting structural features and protein 3D structure. (a) Inter-residue geometrics, including distances ($d$) and orientations, three dihedral ($\omega, \theta_{ij}, \theta_{ij}$) and two planar ($\phi_{ij}, \phi_{ji}$) angles. (b) Outline of the PSP based on structural features from MSA via energy minimization and full-atom relaxation.}
	\label{Rosetta}
\end{figure*}

DESTINI~\citep{MuGao2019DESTINIAD} has two main components: a fully convolutional residual neural network for contact prediction with a template-based structural modelling procedure. Other than the distance map between $C_\beta$ atoms, \cite{JianyiYang2019ImprovedPS} and \cite{JinLi2021StudyOR} predicted orientations between residues via residual networks, which theoretically fully define the relative positions of the backbone atoms of two residues. The protein 3D structure is obtained from these inter-residue geometrics by energy minimization, then refined by full-atom relaxation; the pipeline is shown in Figure~\ref{Rosetta}. AlphaFold~\citep{AndrewWSenior2020ImprovedPS} optimized the potential and constructed predicted distances between pairs of residues through a simple gradient descent algorithm, which can generate structures directly, ignoring exhausting sampling procedures. Structural features-based PSP models commonly have two steps: contact predicting and structure modelling, which have so far been restricted to the accuracy of individual components, even though scientists seek to utilize different information (MSAs, templates, biochemical properties, etc.). In contrast, end-to-end differentiable models have obtained remarkable achievements~\citep{YannLeCun2015DeepL}.
\subsubsection{End-to-end PSP}
RGN~\citep{MohammedAlQuraishi2018EndtoEndDL} predicted torsional angles from protein sequences fed into computational units based on LSTM, which are sequentially translated into Cartesian coordinates to generate the predicted structure. \cite{AlbertCosta2021DistillationOM} harvested individual and contextual residual embeddings produced by MSA Transformer, assigned to nodes and edges to Graph Transformer~\citep{YunshengShi2020MaskedLP}. The birth of the highly-accurate model, AlphaFold2 (AF2)~\citep{JohnMJumper2021HighlyAP} has stimulated the development of end-to-end models for PSP.

\begin{figure}[t]
	\centering
	\subfloat[AF2]
	{
		\label{AF2}
		\includegraphics[width=0.95\linewidth]{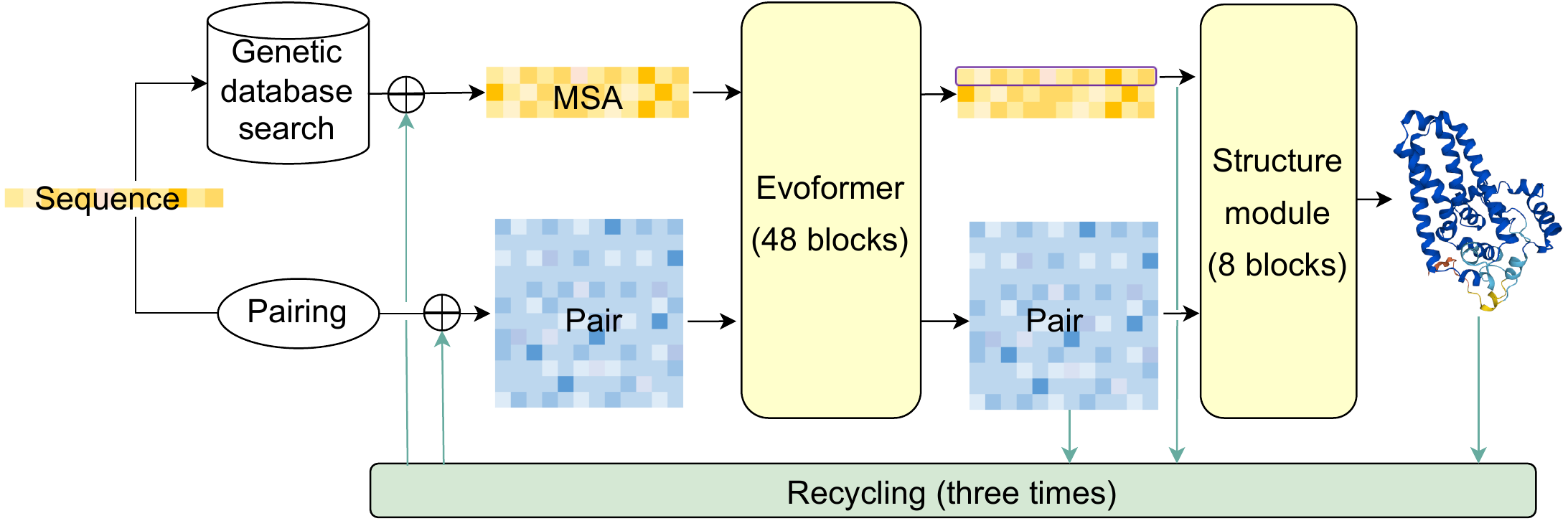}
	}\\
	\subfloat[RosettaFold]
	{
		\label{RosettaFold}
		\includegraphics[width=0.95\linewidth]{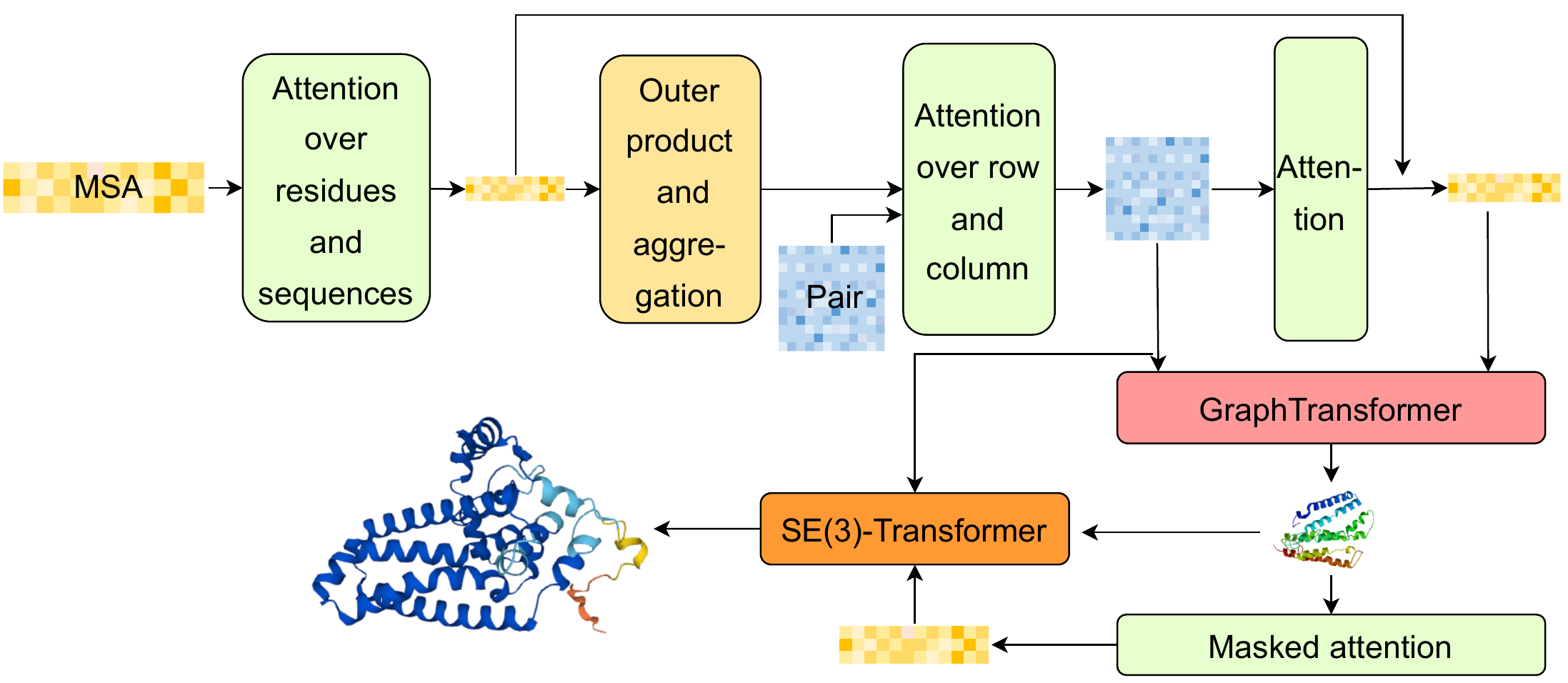}
	}
	\caption{A simplified schematic of AF2 and RosettaFold architectures.}
	\label{AF2 and RosettaFold}
\end{figure}

AF2 can produce even near-experimental results in most cases, whose accuracy was vastly higher than that of other competing methods in CASP14. The overall model architecture of AF2 is exhibited in Figure~\ref{AF2}. In addition to the above-mentioned Evoformer as mentioned earlier (listed in Table~\ref{Table_pLMs}, shown in Figure~\ref{Evoformer}), AF2 also has a decoder, the structure module that has eight layers with shared weights with the pair and first row MSA representations from the Evoformer as input. Each layer updates the single representation and the backbone frames, parameterized as Euclidean transforms. The structure module mainly includes the Invariant Point Attention (IPA) module, which is a form of attention that acts on a set of frames and is invariant under global Euclidean transformations because of the invariant operation, L2-norm of the global transformed vector. Moreover, AF2 executes the network three times with minor extra training time, embedding the previous outputs as additional inputs, making the network more profound and relatively important. There are various models with incremental improvements on different aspects concerning AF2, reducing training time for FastFold~\citep{ShengganCheng2022FastFoldRA} and Uni-Fold~\citep{ZiyaoLi2022UniFoldAO}, faster MSA generation for ColabFold~\citep{mirdita2022colabfold}, denoising the searched MSA or generating virtual MSA for EvoGen~\citep{JunZhang2022FewShotLO} that is useful for proteins lacking sequence homology, and re-implementation of AF2(HelixFold~\citep{GuoxiaWang2022HelixFoldAE}, MEGA-Fold~\citep{SiruiLiu2022PSPMP}, OpenFold~\citep{Ahdritz_OpenFold_2021}).

As for another famous PSP work, RoseTTAFold~\citep{MinkyungBaek2021AccuratePO} (as shown in Figure~\ref{RosettaFold}), it is a three-track model with attention layers in which information flows back and forth at the 1D, 2D, and 3D levels between sequences, distances, and coordinates, which is mainly consisted of seven modules. Firstly, the MSA features are processed by attention over rows and columns, and then the processed features are aggregated by the outer product that can obtain the correlation (coevolution) between two residues in each sequence to update pair features, which are refined via axial attention. Next, the MSA features are also updated based on attention maps derived from pair features, which had a good agreement with the true contact map. Using these learned MSA and pair features as the node and edge embeddings and building a fully-connected graph, Graph Transformer is employed to estimate the initial 3D structure, and new attention maps can be derived from the current structure to update MSA features. Finally, 3D coordinates are refined by SE(3)-Transformer~\citep{FabianBFuchs2020SE3Transformers3R} based on updated MSA and pair features.

Because of the remarkable success of AF2 and RosettaFold, a set of research has emerged, e.g., the protein-peptide binders identification~\citep{LiweiChang2022AlphaFoldET}, applying to small molecules~\citep{MaartenLHekkelman2021AlphaFillET}, antibody structure prediction~\citep{JeffreyARuffolo2022FastAA}, protein complex structure prediction~\citep{RichardEvans2021ProteinCP, PatrickBryant2021ImprovedPO}, RNA 3D structure prediction~\citep{TaoShen2022E2Efold3DED}, and generating multiple protein conformations~\citep{RichardAStein2022SPEACH}, so on and so forth, which are introduced in the following two subsections. A set of representative methods for PSP and related tasks are shown in Table~\ref{Table_PSP}.

\subsubsection{Single Sequence Structure Prediction via pLMs}
\begin{figure*}[h]
	\begin{center}
		\includegraphics[width=0.95\linewidth]{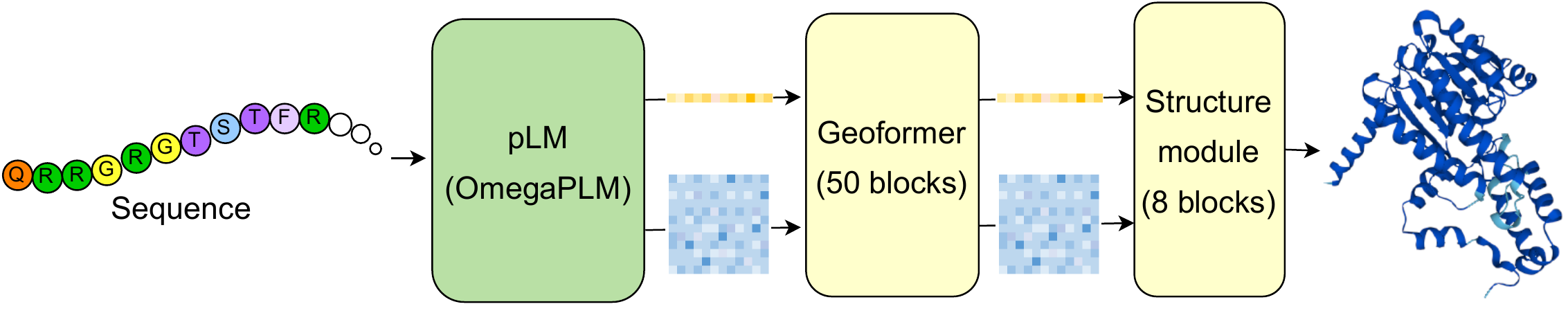}
	\end{center}
	\caption{Model architecture of OmegaFold.}
	\label{OmegaFold}
\end{figure*}
These mainstream PSP pipelines heavily rely on MSAs, which becomes a kind of bottleneck for the reason that it is time-consuming and resource-intensive to search for MSAs and templates. Besides, a protein sequence can theoretically determine its structure; MSA-aware models may memorize the determined structures of similar proteins for PSP, making it difficult for us to understand the mechanism of folding in reality. In order to predict structure from only sequences without MSAs, a set of recent works deemed pLMs can capture grammatically structural information from protein sequence databases. OmegaFold~\citep{RuidongWu2022HighresolutionDN} utilized OmegaPLM to obtain the residue and pairwise embeddings from a single sequence, which are fed into Geoformer. Different from the triangle multiplicative and triangle attention modules applied in Evoformer to enforce the edge representation to satisfy the triangle inequality of distances, Geoformer considers maintaining geometric consistency in the high-dimensional space as well as in the Euclidean space with the help of the structure module implemented by AF2. The model architecture is demonstrated in Figure~\ref{OmegaFold}. The results of contact predictions illustrated that accuracy improves while inconsistency drops with stacked Geoformer layers. Thus, OmegaFold enables accurate predictions on orphan proteins, similar to AF2 structures. Other than OmegaFold, a contemporaneous work, ESMFold~\citep{ZemingLin2022LanguageMO}, has trained large-scale pLMs (ESM2) to learn more structural and functional properties to replace the role of MSAs; similar examples are HelixFold-Single~\citep{XiaominFang2022HelixFoldSingleMP}, RGN2~\citep{RatulChowdhury2021SinglesequencePS}. While trRosettaX-Single~\citep{WenkaiWang2022SinglesequencePS} used pre-trained pLMs (ESM-1b) and retrained it based on supervised learning (s-ESM-1b), distilling knowledge from a pre-trained MSA-based network ($\text{Res2Net\_MSA}$) to the student network ($\text{Res2Net\_Single}$) to predict distance and orientations.
\subsubsection{Diversified Progress in the Protein Field}
\cite{JamesPRoney2022StateoftheArtEO} found that AlphaFold has the ability to learn a highly accurate biophysical energy function and find a low-energy conformation using co-evolutionary information, which is important for single-sequence PSP using physical principles~\citep{roney2022evidence}. However, AF2 structures may contain protein segments that are placed with uncertainty~\citep{akdel2022structural} that need manual inspection instead of purely relying on machines. Most native proteins folded into stable conformations with the lowest free energy~\citep{AnfinsenCb1975ExperimentalAT}. Therefore, both
Atom Transformer~\citep{YilunDu2020EnergybasedMF} and GraphEBM~\citep{DeqinLiu2022SE3EG} are the energy-based models for protein side-chain conformation with different networks, a transformer model and a GNN~\citep{SoumyaSanyal2020ProteinGCNPM, busbridge2019relational} enhanced model, which predicted scores of side-chain conformations for given structures.

Based on AlphaFold-Multimer, ColAttn~\citep{BoChen2022ImproveTP} made use of pLMs to identify interologs of a complex by estimating the column attention weight matrix. \cite{ArneElofsson2022PredictingTS} completed a large protein complex assembly using a Monte Carlo tree search to address the problem of AlphaFold-multimer, whose performance declined rapidly for proteins with three or more chains. Colossal-AI~\citep{bian2021colossal} team has developed pipelines (xTrimoMultimer, xTrimoABFold and xTrimoDock) based on pLMs for the structure prediction of protein complexes and antibodies. To predict the structures of protein-nucleic acid complexes, \cite{MinkyungBaek2022AccuratePO} proposed RoseTTAFoldNA based on RoseTTAFold, which is a unified framework for protein-DNA, protein-RNA complexes, and RNA tertiary structures.

When having a coarse-grained representation of protein structures, EquiFold~\citep{JaeHyeonLee2022EquiFoldPS} predicted protein structures via iterative refinement using an SE(3)-equivariant~\citep{YiLunLiao2022EquiformerEG} neural network. Structure refinement also takes coarse structures as input and output refined coordinates. This process has been integrated into some PSP models, like AF2 and RoseTTaFold. Different from DeepACCNet~\citep{NaozumiHiranuma2020ImprovedPS} and GNNRefine~\citep{XiaoyangJing2021FastAE}, which used predicted distances to guide PyRosetta~\citep{SidharthaChaudhury2010PyRosettaAS} protein structure refinement, a SE(3)-equivariant graph transformer network that is equivariant to the rotation and translation of coordinates was developed in ATOMRefine~\citep{TianqiWu2022AtomicPS} for all-atom structural refinement end-to-end, where initial structures were generated by AF2. In contrast, GNNRefine performs relatively poorly on AF2 structures. Researchers have become adept at working out more challenging problems. For example, DCGAN~\citep{NamrataAnand2018GenerativeMF} has applied generative adversarial networks (GANs) to generate pairwise distance maps from corrupted protein structures and thus recover robust 3D structures.  \cite{KevinEWu2022ProteinSG} trained a denoising diffusion-based generative model with only a vanilla transformer, which generated high-quality, and diverse protein structures inspired by the huge success of diffusion models in a wide range of data modalities~\citep{RobinRombach2022HighResolutionIS, SimonRouard2021CRASHRA}, such kind of generative models can benefit structure refinement, protein conformation and structure prediction by generating biologically plausible and robust structures, it is worth to expect more successes achieved by diffusion model in the biomedical field.

\begin{table*}[t]
	\caption{Representative methods for PSP and related tasks.}
	\label{Table_PSP}
	\setlength{\tabcolsep}{2.8pt}
	\centering
	\small
	\begin{adjustbox}{max width=\linewidth}
		\begin{threeparttable} 
			\begin{tabular}{llllllll}
				\toprule 
				Model &  Evolution  & Energy  & Main Network & Task \\
				\hline 
				\href{https://github.com/aqlaboratory/rgn}{RGN}~\citep{MohammedAlQuraishi2018EndtoEndDL} & \Checkmark & \XSolidBrush & LSTM & PSP \\
				\multirow{2}{*}{\cite{AlbertCosta2021DistillationOM}} & \multirow{2}{*}{\Checkmark} & \multirow{2}{*}{\XSolidBrush} & MSA Transformer & \multirow{2}{*}{PSP}\\
				& & & Graph Transformer~\citep{YunshengShi2020MaskedLP} & \\
				
				\href{https://github.com/deepmind/alphafold}{AF2}~\citep{JohnMJumper2021HighlyAP} & \Checkmark & \XSolidBrush & Evoformer, IPA & PSP \\
				\multirow{2}{*}{\href{https://github.com/RosettaCommons/RoseTTAFold}{RoseTTAFold}~\citep{MinkyungBaek2021AccuratePO}} & \multirow{2}{*}{\Checkmark} & \multirow{2}{*}{\XSolidBrush} &  Graph Transformer & \multirow{2}{*}{PSP} \\
				& & & SE(3)-Transformer~\citep{FabianBFuchs2020SE3Transformers3R} & \\
				
				\href{https://github.com/hpcaitech/FastFold}{FastFold}~\citep{ShengganCheng2022FastFoldRA} & \Checkmark & \XSolidBrush & AF2 & PSP \\
				\href{https://github.com/sokrypton/ColabFold}{ColabFold}~\citep{mirdita2022colabfold} & \Checkmark & \XSolidBrush & AF2 & PSP \\
				\href{https://github.com/PaddlePaddle/PaddleHelix/tree/dev/apps/protein_folding/helixfold}{HelixFold}~\citep{GuoxiaWang2022HelixFoldAE} & \Checkmark & \XSolidBrush & AF2 & PSP \\
				\href{https://gitee.com/mindspore/mindscience/tree/master/MindSPONGE/applications/MEGAProtein}{MEGA-Fold}~\citep{SiruiLiu2022PSPMP} & \Checkmark & \XSolidBrush & AF2 & PSP \\
				\href{https://github.com/dptech-corp/Uni-Fold}{Uni-Fold}~\citep{ZiyaoLi2022UniFoldAO} & \Checkmark & \XSolidBrush & AF2 & PSP \\
				
				EvoGen~\citep{JunZhang2022FewShotLO} & \Checkmark & \XSolidBrush & Attention & MSA Generation \\
				
				\href{https://github.com/facebookresearch/esm}{ESMFold}~\citep{ZemingLin2022LanguageMO} & \XSolidBrush & \XSolidBrush &ESM-2 & Single Sequence PSP \\  
				\href{https://github.com/HeliXonProtein/OmegaFold}{OmegaFold}~\citep{RuidongWu2022HighresolutionDN} & \XSolidBrush & \XSolidBrush & OmegaPLM, Geoformer & Single Sequence PSP \\
				\href{https://github.com/PaddlePaddle/PaddleHelix/tree/dev/apps/protein_folding/helixfold-single}{HelixFold-Single}~\citep{XiaominFang2022HelixFoldSingleMP} & \XSolidBrush & \XSolidBrush & AF2 & Single Sequence PSP\\
				\href{https://github.com/aqlaboratory/rgn2}{RGN2}~\citep{RatulChowdhury2021SinglesequencePS} & \XSolidBrush & \XSolidBrush & AminoBERT & Single Sequence PSP \\
				
				trRosettaX-Single~\citep{WenkaiWang2022SinglesequencePS} & \XSolidBrush & \Checkmark & s-ESM-1b, $\text{Res2Net\_Single}$ & Single Sequence PSP\\
				EquiFold~\citep{JaeHyeonLee2022EquiFoldPS} & \XSolidBrush & \XSolidBrush & SE(3)-Transformer & Antibody Structure Prediciton \\
				
				\href{https://github.com/Graylab/IgFold}{IgFold}~\citep{JeffreyARuffolo2022FastAA} & \XSolidBrush & \XSolidBrush & Graph Transformer, AntiBERTy, IPA & Antibody Structure Prediciton \\

				\href{https://github.com/deepmind/alphafold}{AlphaFold-Multimer}~\citep{RichardEvans2021ProteinCP} & \Checkmark & \XSolidBrush & AF2 & Complex Structure Prediciton \\
				\cite{PatrickBryant2021ImprovedPO} & \Checkmark & \XSolidBrush & AF2 & Complex Structure Prediciton\\
				
				ColAttn~\citep{BoChen2022ImproveTP}	& \Checkmark & \XSolidBrush & MSA Transformer, AlphaFold-Multimer & Complex Structure Prediciton\\
				
				\cite{ArneElofsson2022PredictingTS} & \Checkmark & \XSolidBrush & AlphaFold-Multimer & Complex Structure Prediciton \\
				
				\multirow{2}{*}{\href{https://github.com/uw-ipd/RoseTTAFold2NA}{RoseTTAFoldNA}~\citep{MinkyungBaek2022AccuratePO}} & \multirow{2}{*}{\Checkmark} & \multirow{2}{*}{\XSolidBrush} & \multirow{2}{*}{RoseTTAFold} & Structure Prediction of RNA, Protein \\
				& & & &-DNA $\&$ Protein-RNA Complexes \\
				\href{https://github.com/RFOLD/RhoFold}{E2Efold-3D}~\citep{TaoShen2022E2Efold3DED} &  \checkmark & \XSolidBrush & Transformer, IPA & RNA Structure Prediction \\
				
				DCGAN~\citep{NamrataAnand2018GenerativeMF} & \XSolidBrush & \checkmark & GAN & Structure Generation\\
				\cite{KevinEWu2022ProteinSG} & \XSolidBrush & \XSolidBrush & Transformer & Structure Generation\\
				
				\href{https://github.com/facebookresearch/protein-ebm}{Atom Transformer}~\citep{YilunDu2020EnergybasedMF} & \XSolidBrush & \Checkmark & Transformer & Protein Conformation\\
				\href{https://github.com/RSvan/SPEACH_AF}{$\text{SPEACH\_AF}$}~\citep{RichardAStein2022SPEACH} & \Checkmark & \XSolidBrush & AF2 & Protein Conformation \\
				GraphEBM~\citep{DeqinLiu2022SE3EG} & \XSolidBrush & \Checkmark & DimeNet~\citep{JohannesKlicpera2020DirectionalMP} & Protein Conformation \\
				
				\href{https://codeocean.com/capsule/5769140/tree/v1}{GNNRefine}~\citep{XiaoyangJing2021FastAE} & \XSolidBrush & \XSolidBrush & GNN & Structure Refinement \\
				ATOMRefine~\citep{TianqiWu2022AtomicPS} & \XSolidBrush & \XSolidBrush & SE(3) Graph Transformer & Structure Refinement \\
				
				\toprule
			\end{tabular}
			\begin{tablenotes} 
				\item \Checkmark and \XSolidBrush appeared in the column of Evolution, and Energy represents this model whether used this information or not. The model name with colour is linked with codes or its server page. Most of these present methods are pLM-aware models. $\&$: and.
			\end{tablenotes} 
		\end{threeparttable}
	\end{adjustbox}
\end{table*}

\section{Discussion: Limitations and Future Trends}
With the million-level protein sequence databases, pLMs have been becoming larger and larger (billion-level parameters for ESM-2), which are dominated by big companies in reality. For example, DeepMind used 128 TPU v3 cores to train and fine-tune AF2 over one week. Training such deep learning models might only be accessible to large companies such as Google; it seems hard for academic research groups to learn protein embeddings from the start, which is also a burden for the theme of a green environment. Considering this problem, researchers can put more attention to methods' innovations. 

Firstly, why not better utilize these large-scale pre-trained pLMs? How to best leverage them is still under-explored compared to pre-trained language models in NLP; it means developing suitable and special transfer learning methods to adapt knowledge to various downstream tasks. Knowledge distillation is inspiring; MSA, structure and function information can be distilled and transferred~\citep{AlbertCosta2021DistillationOM}, which is used in NLP~\citep{ZiqingYang2020TextBrewerAO}. Secondly, a large-scale pLM cannot tackle all problems, e.g., some reports have indicated that AF2 does not appear to be well suited to predict the impact of mutations on proteins~\citep{MarinaAPak2021UsingAT, GwenRBuel2022CanAP}. Structure-triggered tasks cannot be directly transferred to other prediction tasks~\citep{MingyangHu2022ExploringE}. Furthermore, \cite{ErikNijkamp2022ProGen2ET} found bigger models may not translate into better zero-shot fitness performance. Therefore, there still exists space for different models to tackle various problems. Thirdly, ESM-2 concluded that the improvement of low-scale models is easily saturated with high evolutionary depth, while with the low evolutionary depth, it continues when models' sizes increase, which illustrates that combining extra information suitably, including injecting biological and physical priors, using evolutionary information, different levels of structures, GO, etc., can reduce models' size and boost their performance. These different types of information (heterogeneous data) are involved with data mining and task design. Thus, multi-task or multi-modal learning is a direction that deserves to have more exploration~\citep{TristanBepler2021LearningTP}. Finally, since sequence mutations can cause genetic diseases, their effects on function form a landscape that reveals how function constrains sequence. Predicting robust protein structures and understanding the functional effects of sequence mutations are critical. 

The architectures of most current pLMs are the same as LMs' in NLP, i.e., people use these LMs in the protein communities without much adaptation and modification, which hinders the development of pLMs and the understanding of protein grammars. Techniques present in LMs are also used in pLMs, like masking strategies. Proteins and Languages have similarities, but they are not the same as we have mentioned in Section~\ref{Protein and Language}. Therefore, designing proper pLMs and developing suitable methods for protein data considering its properties is an important and urgent problem. 

Despite the proliferation of LMs, these pLM-based models still lack interpretability, hindering people from understanding the mechanism behind protein folding. AF2-like models cannot provide the detailed understanding of molecular and chemical interactions that is important for studies of molecular mechanisms and structure-based drug design. Thus, model interpretability helps researchers find protein grammars useful in biomedical applications. Visualization is a tool that can record the process of protein folding or how the protein functions during various activities. Therefore, developing a high-quality protein tokenization method or combining it with other machine learning methods is appealing, like Bayesian modelling, Monte Carlo strategy, energy functioning, Markov random field, direct coupling analysis~\citep{MartinWeigt2009IdentificationOD}, or diffusion models. 

Due to the different large databases of proteins, people need to be concerned about the wrong information in the databases. Some research groups choose to build new datasets to satisfy their unique needs; data leakage, bias, and ethics should be considered in this situation. For instance, ProteinKG25 was created for utilizing protein sequence and functions~\citep{NingyuZhang2022OntoProteinPP} better to get meaningful embeddings. On the other hand, building complete, reliable, and just benchmarks is essential to evaluating various models and promoting the appearance of solid methods.

Scientists have been developing a unified model for DNA, RNA, and protein engineering in the biomedical domain~\citep{BenyouWang2022PretrainedLM}, BERT-RBP~\citep{KeisukeYamada2021PredictionOR} adopted BERT architecture to predict RNA-protein interactions, still has a distance to go to interpret their relationships well. Single protein sequences, protein-ligand complexes, and RNA complexes structure prediction and conformation modelling have been appealing to a group of researchers. These problems are more complicated than those in PSP, and they are not yet solved. Even for PSP, there is still space for models predicting structures with ultra-high accuracy (resolution less than $0.5\AA$).

\section{Databases}
\label{Databases}
We follow the mapping for LMs from NLP to proteins in ProtTrans~\citep{AhmedElnaggar2022ProtTransTC}, which interpret proteins as "sentences". Therefore, the number of proteins that needed disk space for datasets and official web pages are listed in Table~\ref{datasets}. 

In detail, UniProt~\citep{2012UpdateOA} provides a comprehensive, high-quality, and freely accessible resource of protein information, where UniProt databases include UniProtKB, UniRef~\citep{BarisESuzek2015UniRefCA}, UniPac databases. The UniProt Knowledgebase (UniProtKB) is comprised of two sections, UniProtKB/Swiss-Prot and UniProtKB/TrEMBL. The former provides the reviewed relevant information about a particular protein, including protein and gene names, function, protein-protein interactions, etc. Proteins in the latter largely have no experimental data available, while these unreviewed records are enriched with automatic annotation and classification. The UniRef databases (UniProt Reference Clusters) ~\citep{BarisESuzek2015UniRefCA} provide clustered sets of sequences from the UniProt Knowledgebase (UniProtKB) and provide complete coverage of sequence space at several resolutions (100$\%$, 90$\%$, 50$\%$ identity) in a hierarchical fashion. UniRef50 clusters are generated using UniRef90 clusters, which are based on UniRef100 clusters. UniRef databases have a broad range of usages, like functional annotation, family classification, systems biology, structural genomics, phylogenetic analysis, and mass spectrometry. The UniProt Archive (UniParc) is the most comprehensive publicly accessible non-redundant protein sequence database without record annotation.

The Protein Data Bank (PDB)~\citep{wwpdb2019protein} is a database for the experimentally determined three-dimensional structural data of biological macromolecules, including the sequence and 3D structures of proteins. The number of structures in the PDB has grown at an approximately exponential rate based on historical experience, and there are 170k multi-chain protein structure files as of October 2021. However, a million-level protein structure prediction named as PSP~\citep{SiruiLiu2022PSPMP} was presented based on PDB after the appearance of AF2 with 570k true structure sequences and 745k complementary distillation sequences. 

The CATH~\citep{ChristineAOrengo1997CATHA} is a novel public hierarchical classification database of protein domain structures for public download, where these domains are obtained from protein three-dimensional structures deposited in the PDB. The domains are classified into four main levels: protein class (C), which describes the secondary structure composition of each domain, i.e., all alpha, all beta, mixed alpha-beta, or few secondary structures; at the architecture (A) level, the arrangement of secondary structures is summarized; at the topology/fold (T) level, the sequential connectivity is taken into account; at the homologous superfamily (H) level, proteins have a demonstrable evolutionary relationship. Structured Transformer~\citep{JohnIngraham2019GenerativeMF} uses CATH4.2, processed and cluster-split into training, validation, and test sets, which contain 18025 chains in the training set, 1637 chains in the validation set, and 1911 chains in the test set. The latest release of CATH-Gene3D (v4.3) has 151 million protein domains classified into 5,841 superfamilies and is to predict the locations of structural domains on millions of publicly available protein sequences. 

The Big Fantastic Database (BFD) is a sequence profile database and is one of the largest metagenomic databases, as it keeps one copy of duplicates from UniProt and other big protein databases. These sequences were clustered by a sequence identity cut-off of 30$\%$ and a coverage threshold of 90$\%$ using MMseqs2/Linclust~\citep{MartinSteinegger2018ClusteringHP, MartinSteinegger2019ProteinlevelAI, HOU2022434}, e.g., the BFD30 dataset was clustered to 65 million clusters at 30$\%$ sequence identity. Searching for a protein against BFD is slow but more sensitive. Sometimes it is convenient to use MSA, computed from each of the BFD clusters.

Pfam~\citep{SaraElGebali2019ThePP} is a database of protein families used extensively in bioinformatics. It aims to provide a complete and accurate classification of protein families and domains, which includes annotations and alignments generated by hidden Markov models (HMMs). \textit{pfamseq} is a profile HMM that is queried against a sequence database based on UniProtKB and used to find homologues for a Pfam entry. The recent version of Pfam contains 19,632 families; each family includes descriptions, alignments, architectures, etc.

The AlphaFold Protein Structure Database (AlphaFoldDB) is a collection of protein structure predictions created by DeepMind in partnership with EMBL-EBI. These 3D structures are predicted by AlphaFold, which achieves accuracy competitive with experiments. The latest database release contains over 200 million entries for UniProt.

ProteinKG25 is a large-scale Knowledge Graph (KG) dataset with aligned descriptions and protein sequences, respectively, to Gene Ontology (GO) terms and protein entities~\citep{NingyuZhang2022OntoProteinPP}. Go terms are seen as graph nodes, and the relationships between the terms are edges. ProteinKG25 combines these two different structures, GO and Gene Annotation into a unified KG for training LMS that incorporates GO information. It includes most of the triplets (Protein-GO triplets and GO-GO triplets) and a few entities (proteins and gene terms) and relations.

Uniclust databases~\citep{MilotMirdita2017UniclustDO} cluster UniProtKB sequences at the level of 90$\%$, 50$\%$ and 30$\%$ pairwise sequence identity, formed Uniclust90, Uniclust50, Uniclust30. Compared with UniRef, which relied on the CD-HIT software for clustering, Uniclust databases used the developed software suite MMseqs2, which had higher consistency in the functional annotation. Furthermore, Uniclust sequences are annotated with matches to other commonly used datasets (Pfam, SCOP~\citep{TimHubbard1997SCOPAS} and PDB), which may not be annotated in UniProt with HHblits~\citep{MartinSteinegger2019HHsuite3FF}.

The Structural Classification of Proteins (SCOP) database~\citep{TimHubbard1997SCOPAS} was created by manual classification of protein structural domains based on similarities of their structures and evolutionary relationships. Similar to CATH and Pfam, the original hierarchical organizations of SCOP are class, fold, superfamily, and family. The new version, SCOP2, was released in 2020 to provide a better database for protein structure annotation and classification. Structural Classification of Proteins—extended (SCOPe) extends the SCOP database of protein structural relationships. The ASTRAL compendium provides tools to help research protein structure and evolution. The SCOPe has corrected errors in SCOP and incorporated the Astral database~\citep{JohnMarcChandonia2017SCOPeMC}. The latest release of SCOPe ASTRAL 2.08 has more than 50K protein sequences with known structures and SCOP classifications.

Critical Assessment of Protein Structure Prediction (CASP) is a worldwide experiment for PSP that will have been conducted 15 times by the end of 2022. Research groups from all over the world participate in CASP to objectively test their structure prediction methods. By categorizing different themes, like quality assessment, model refinement, domain boundary prediction, protein complex structure prediction, etc., and selecting target proteins, CASP researchers can identify what progress has been made and highlight the future efforts that may be most productively focused on.

\begin{table*}[t]
	\caption{Information of Protein Databases}
	\label{datasets}
	\setlength{\tabcolsep}{2.8pt}
	\centering
	\small
	\begin{adjustbox}{max width=\linewidth}
		\begin{threeparttable} 
			\begin{tabular}{llllllll}
				\toprule 
				Dataset & $\#$Proteins & Disk Space & Description & Link \\
				\hline 
				UniProtKB/Swiss-Prot & 500K & 0.59GB& knowledgebase & \url{https://www.uniprot.org/uniprotkb?query=*} \\
				UniProtKB/TrEMBL & 229M & 146GB & knowledgebase & \url{https://www.uniprot.org/uniprotkb?query=*} \\
				UniRef100 & 314M &76.9GB & clustered sets of sequences & \url{https://www.uniprot.org/uniref?query=*}  \\
				UniRef90 & 150M  & 34GB & 90$\%$ identity & \url{https://www.uniprot.org/uniref?query=*}\\
				UniRef50 & 53M &  10.3GB & 50$\%$ identity & \url{https://www.uniprot.org/uniref?query=*}\\
				UniParc & 528M  & 106GB & Sequence & \url{https://www.uniprot.org/uniparc?query=*}\\
				PDB & 190K & 50GB & 3D structure & \url{https://www.wwpdb.org/ftp/pdb-ftp-sites}\\
				CATH4.3 & N/A & 1073MB & hierarchical classification & \url{https://www.cathdb.info/} \\
				BFD & 2500M & 272GB & sequence profile & \url{https://bfd.mmseqs.com/} \\
				Pfam & 47M & 14.1GB & protein families &\url{https://www.ebi.ac.uk/interpro/entry/pfam/} \\
				AlphaFoldDB & 214M & 23 TB & predicted 3D structures &\url{https://alphafold.ebi.ac.uk/} \\
				ProteinKG25 & 5.6M & 147MB & a KG dataset with GO & \url{https://drive.google.com/file/d/1iTC2-zbvYZCDhWM_wxRufCvV6vvPk8HR} \\
				Uniclust30 & N/A & 6.6GB & clustered protein sequences & \url{https://uniclust.mmseqs.com/} \\
				SCOP & N/A & N/A & structural classification & \url{http://scop.mrc-lmb.cam.ac.uk/} \\
				SCOPe & N/A & 86MB & extented version of SCOP & \url{http://scop.berkeley.edu}\\
				
				\toprule
			\end{tabular}
			\begin{tablenotes} 
				\item K, thousand; M, million, disk space is in GB or TB (compressed storage as text), which is estimated data influenced by the compressed format.
			\end{tablenotes} 
		\end{threeparttable}
	\end{adjustbox}
\end{table*}

\section{Conclusion}
This paper systematically summarizes recent advances in pLMs, PSP, and related tasks, including background, why and how pLMs are used in protein representation learning, existing pLMs and PSP methods, the relationships between pLMs and PSP, and how pLMs function in the development of protein folding. A set of databases are introduced. Furthermore, we also discuss some limitations, possible tackling directions, and future trends. Finally, we expect that LMs and pLMs can aid more in the specific biochemical, biomedical, and bioinformatic domains.

\bibliographystyle{apalike}
\bibliography{ref}
\end{document}